\documentclass[tran,twocolumn,nonote,noversion,halfsqueezed]{cdcarticle}

\def\MODE{2}

\if\MODE1\else\usepackage{amsthm}\fi

\usepackage{amsfonts,amssymb,amsmath,mathrsfs}
\usepackage{mathtools,enumerate,colonequals}
\usepackage{graphicx}
\if\MODE1\usepackage{natbib}\fi
\usepackage{subcaption}
\usepackage[hidelinks]{hyperref}

\usepackage{pgf}
\usepackage{tikz}
\usetikzlibrary{matrix,fit,backgrounds,plotmarks,calc}

\usepackage{pgfplots}
\pgfplotsset{width=9.1cm,height=5.8cm,compat=newest}
\pgfplotscreateplotcyclelist{plain black}{%
	every mark/.append style={fill=gray,scale=0.6},mark=*\\%
}

\tikzset{
dashedx/.style={}
}

\def\qed{\hfill ~\rule[0pt]{5pt}{5pt}\par\medskip}		

\if\MODE1\else
 \newtheorem{thm}{Theorem}
 \newtheorem{lem}[thm]{Lemma}
 \newtheorem{prop}[thm]{Proposition}
 \newtheorem{rem}[thm]{Remark}
 \newtheorem{defn}[thm]{Definition}
 \newtheorem{cor}[thm]{Corollary}
\theoremstyle{definition}
 \newtheorem{exmp}{Example}
 \newtheorem{prob}{Problem}
 \newenvironment{pf}{{\noindent\bf Proof.}}{ \hfill ~\qed}
\fi

\def\tp{\mathsf{T}}

\newcommand{\defeq}{\colonequals}

\newcommand{\nodelay}{\xrightarrow{0}}
\newcommand{\delay}{\xrightarrow{1}}

\DeclareMathOperator{\ee}{\mathbb{E}}
\DeclareMathOperator{\Trace}{\mathbf{trace}}

\DeclareMathOperator{\Span}{\mathbf{lin}}

\newcommand{\bmat}[1]{\begin{bmatrix}#1\end{bmatrix}}

\renewcommand{\phi}{\varphi}

\newcommand{\Lab}{\mathcal{L}}
\newcommand{\Info}{\mathcal{I}}
\newcommand{\Hist}{\mathcal{H}}

\newcommand{\Msg}{\mathcal{M}}
\newcommand{\Mem}{\mathcal{R}}

\if\MODE1\else

	\def\note#1{}
\fi

\begin{document}

\if\MODE1\begin{frontmatter}\fi

\title{Optimal Decentralized State-Feedback Control \\with Sparsity and Delays} 

\if\MODE1
	\author[First]{Andrew~Lamperski}\ead{\\alampers@umn.edu},
	\author[Second]{Laurent~Lessard}\ead{lessard@berkeley.edu}
	\address[First]{Department of Electrical and Computer Engineering, 
          University of Minnesota, Minneapolis, MN 55455, USA.}
	\address[Second]{Department of Mechanical Engineering,
		University of California Berkeley, Berkeley, CA 94720, USA.}
\else
	\author{Andrew~Lamperski~\and~Laurent~Lessard}
\fi
\if\MODE1\else\note{Accepted to Automatica}\fi

\maketitle

\begin{abstract}
This work presents the solution to a class of decentralized linear quadratic state-feedback control problems, in which the plant and controller must satisfy the same combination of delay and sparsity constraints. Using a novel decomposition of the noise history, the control problem is split into independent subproblems that are solved using dynamic programming. The approach presented herein both unifies and generalizes many existing results.
\end{abstract}

\if\MODE1\end{frontmatter}\fi

\section{Introduction}\label{sec:intro}

While optimal decentralized controller synthesis is difficult in general~\cite{tsitsikliscomplexity1984,witsenhausencounterexample1968}, much progress has been made toward identifying tractable subclasses of problems. Two closely related conditions, partial nestedness and quadratic invariance, guarantee respectively that the optimal solution for an LQG control problem is linear~\cite{hochu}, and that optimal synthesis can be cast as convex program~\cite{qimurti04,rotkowitz06}.
These results alone do not guarantee that the optimal controller can
be efficiently computed since the associated optimization problems
are large.

More efficient computational tools have been developed for linear quadratic problems. In particular, linear matrix inequalities have been used to solve the state-feedback~\cite{rantzer06} and output-feedback~\cite{gattamigeneralized2010,rantzerseparation2006} cases. Output-feedback problems with delays have also been solved using spectral factorization and quadratic programming~\cite{lamperskioutput2013}.
A drawback of purely computational approaches is that little insight is gained into the structure of optimal controllers. 
However, explicit solutions that provide efficient computation in addition to a physical interpretation for the states of the controller have been found separately for the delay and sparsity cases.

\if\MODE1\textbf{Delay case: }\else\paragraph{Delay case:}\fi
All controllers eventually measure the global state, but not
necessarily simultaneously. Instances with a one-timestep delay
between controllers were solved by extending classical dynamic
programming
arguments~\cite{kurtaranlinearquadraticgaussian1974,sandellsolution1974,yoshikawadynamic1975}. In the linear quadratic setting, the state-feedback problem with delays characterized by a graph is solved in~\cite{lamperskidynamic2012}.
The optimal policy is dynamic, with controller states formed by
delayed estimates of the global state.

\if\MODE1\textbf{Sparsity case: }\else\paragraph{Sparsity case:}\fi
All state measurements are transmitted instantly, but not all controllers receive all measurements. This is equivalent to a
delayed system where each delay is either zero or infinite. 
Explicit solutions for a two-controller system have been solved using a spectral
factorization approach~\cite{swigartexplicit2010a} and dynamic
programming~\cite{swigartexplicit2010b}. The spectral factorization
method was extended to a  general class of quadratically
invariant sparsity
patterns~\cite{shah10,swigart_thesis}. 
Again, the controller states can be interpreted as estimates of the global states conditioned on the particular subsets of the available information.

This paper unifies the treatment of state feedback with
sparsity constraints, \cite{shah10,swigart_thesis},  and delay
constraints, \cite{lamperskidynamic2012}, by 
considering an information flow characterized by a directed
graph. Each edge may be labeled with a~$0$ for instantaneous
information transfer or with a~$1$ for a one-timestep delay. See below for an example of such a graph. The $0$--$1$ convention is merely for ease of exposition; the case of general inter-node delays is discussed in Section~\ref{sec:prob_statement}. 

\begin{exmp}\label{ex:threeNode}
Consider the network graph of Fig.~\ref{fig:ex1_diagram}.
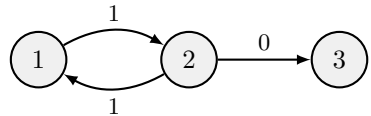
\begin{figure}[bht]
\centering
\begin{tikzpicture}[thick,node distance=2.0cm,>=latex]
	\def\fs{\small}  
	\tikzstyle{block}=[circle,draw,fill=black!6,minimum height=2.1em]
	\tikzstyle{loopstyle}=[loop,looseness=8]
	\node [block](N1){$1$};
	\node [block, right of=N1](N2){$2$};
	\node [block, right of=N2](N3){$3$};
	\draw [->] (N1) edge [out=30, in=150]  node[above]{\fs$1$} (N2);
	\draw [<-] (N1) edge [out=-30,in=-150] node[below]{\fs$1$} (N2);
	\draw [->] (N2) edge node[above]{\fs$0$} (N3);
\end{tikzpicture}
\caption{Network graph for Example~\ref{ex:threeNode}. Each node represents a subsystem, and the edge labels indicate the propagation delay from one subsystem to another.\label{fig:ex1_diagram}}
\end{figure}
\end{exmp}
Although the example of Fig.~\ref{fig:ex1_diagram} appears simple, it contains both salient features previously discussed: delay constraints (between nodes~1 and~2) and sparsity constraints (between nodes~2 and~3). We will begin in Section~\ref{sec:ex1sol} by sketching our approach as it applies to the simple Example~\ref{ex:threeNode}. In Sections~\ref{sec:prob_statement} and~\ref{sec:main}, we broaden our scope and treat general directed graphs with any number of nodes and an arbitrary but fixed delay pattern.

A fundamental assumption in this work is that the control policies are jointly optimized in order to minimize a global cost function. In our search for the optimal policies, we assume global knowledge of the graph topology, system dynamics, and cost function for the entire time horizon. Similar assumptions are required in the classical LQR problem. For instance, if the plant has time-varying dynamics, the optimal control gain at time $t$ will generally depend on \emph{future} plant parameters.

In other words, the system is decentralized in the sense that controllers have limited state information at run time. However, the \emph{design} of the controllers assumes global knowledge. In the absence of such an assumption,
or when memory is constrained, the resulting problem is nonconvex and very difficult in general~\cite{witsenhausencounterexample1968}. Thus, work on multi-agent control with limited system knowledge typically does not study optimal control~\cite{caooverview2013}, or finds locally optimal solutions to non-convex problems \cite{farokhioptimal2013}. 

In the remainder of the paper, we discuss how our work unifies
existing results in Section~\ref{sec:discussion1} and we discuss its
limitations in Section~\ref{sec:discussion2}. We provide an
illustrative numerical example in Section~\ref{sec:example} and prove the main results in Section~\ref{sec:mainproof}. Finally, we conclude in Section~\ref{sec:conclusion}.

A preliminary version of this work appeared in the conference
paper~\cite{lamperskioptimal2012}. 
The present work includes complete proofs to all results, new illustrative examples, and expanded discussions in Sections~\ref{sec:discussion1}
and~\ref{sec:discussion2}. We also have a new result, Theorem~\ref{thm:msg}, which gives a distributed message passing implementation of the optimal controller.

\section{Solution to Example~\ref{ex:threeNode}}\label{sec:ex1sol}

We begin by sketching our approach for the simple example from Fig.~\ref{fig:ex1_diagram}. The graph indicates
constraints both on information sharing amongst controllers as well as on the system dynamics. In this case, the dynamics are given by discrete-time state-space equations of the form
\begin{multline}\label{ex1_sseqn}
\bmat{x_{t+1}^1\\x_{t+1}^2\\x_{t+1}^3}
=
\bmat{A^{11}_t & A^{12}_t & 0 \\
      A^{21}_t & A^{22}_t & 0 \\
      A^{31}_t & A^{32}_t & A^{33}_t}
\bmat{x_t^1\\x_t^2\\x_t^3} \\
+\bmat{B^{11}_t & B^{12}_t & 0 \\
        B^{21}_t & B^{22}_t & 0 \\
        B^{31}_t & B^{32}_t & B^{33}_t}
\bmat{u_t^1\\u_t^2\\u_t^3} +
\bmat{w_t^1\\w_t^2\\w_t^3}
\end{multline}
for $t = 0,1,\dots,T-1$. The state, input, and disturbance are denoted by $x_t$, $u_t$, and $w_t$ respectively. Each vector is partitioned into subvectors associated with the nodes of the graph. For example, $x^2_t$ is associated with node 2. The dynamics are constrained according to the directed graph. If node $i$ cannot affect node $j$ after a delay of $0$ or $1$, then $A^{ji}_t=0$ and $B^{ji}_t=0$ for all $t$.

We assume that for $i\in\{1,2,3\}$, the initial state and the disturbance vectors $\{x_0^i,w_0^i,\dots,w^i_{T-1}\}$ are independent Gaussian random vectors with means and covariances
\begin{equation}\label{ass:norm_distributions}
x_0^i \sim \mathcal{N}(0,\Sigma_0^i)
\quad
\text{and}
\quad
w_t^i \sim \mathcal{N}(0,W_t^i)
\quad
\text{for all } t.
\end{equation}
The policies of the decision-makers choosing $u_t^i$ are again constrained according to the graph. In particular,
\begin{subequations}\label{ex1_constraints}
\begin{align}
u^1_t &= \gamma_t^1\bigl( x_{0:t}^1,\, x_{0:t-1}^2 \bigr) \\
u^2_t &= \gamma_t^2\bigl( x_{0:t-1}^1,\, x_{0:t}^2 \bigr) \\
u^3_t &= \gamma_t^3\bigl( x_{0:t-1}^1,\, x_{0:t}^2,\, x_{0:t}^3 \bigr)
\end{align}
\end{subequations}
for all $t$, where each $\gamma_t^i$ is a measurable function of the state information that has had sufficient time to propagate to node~$i$. We use the notation $x^i_{0:t}$ to denote the state history $(x^i_0,\dots,x^i_t)$. 

The objective is to choose the policies $\gamma$ that minimize the expected finite-horizon quadratic cost
\begin{equation}\label{eq:cost_function}
\min_\gamma 
\ee^\gamma \Biggl(
\sum_{t=0}^{T-1} \bmat{x_t\\u_t}^\tp \bmat{Q_t & S_t \\ S_t^\tp & R_t} \bmat{x_t\\u_t}
+ x_T^\tp Q_f x_T^{\vphantom{\tp}}
\Biggr)
\end{equation}
where the expectation is taken with respect to the joint probability measure on $(x_{0:T},u_{0:T-1})$ induced by the choice of $\gamma$. We make the standard assumptions that
\begin{align}\label{ass:QRS_assumptions}
\bmat{Q_t & S_t \\ S_t^\tp & R_t}\ge 0,
&&
R_t > 0,
&&
Q_f\ge 0.
\end{align}
We assume that all decision-makers know the underlying network
graph~$G(\mathcal{V},\mathcal{E})$ and all system
parameters~$A_{0:T-1}$, $B_{0:T-1}$, $Q_{0:T-1}$, $R_{0:T-1}$,
$S_{0:T-1}$, and $Q_f$. 
Note that system matrix sizes may also vary with time.

Under the above assumptions, the problem is \emph{partially nested}. Thus, the results from \cite{hochu} imply that the optimal policies $\gamma$ are linear functions.

The main insight behind the method is a reparametrization of the information
constraints that leads to decoupled optimization problems. The
resulting solutions are combined to optimally solve the
original  problem. 

\subsection{Disturbance-feedback representation}\label{sec:ex1_disturbance}

The first step in the solution reparameterizes the input as functions
the initial conditions and the disturbances. As in previous work
on decentralized control,
\cite{gattamigeneralized2010,lamperski_delayed,rantzer06,shah10,swigart_thesis,swigartexplicit2010b},
such a representation enables us to use statistical independence of
the noise terms to simplify derivations.  

Defining $w_{-1} \defeq x_0$, the controllers~\eqref{ex1_constraints} may be equivalently written as
\begin{subequations}\label{ex1_noiseform}
\begin{align}
u^1_t &= \hat\gamma_t^1\bigl( w_{-1:t-1}^1,\, w_{-1:t-2}^2 \bigr) \\
u^2_t &= \hat\gamma_t^2\bigl( w_{-1:t-2}^1,\, w_{-1:t-1}^2 \bigr) \\
u^3_t &= \hat\gamma_t^3\bigl( w_{-1:t-2}^1,\, w_{-1:t-1}^2,\,
w_{-1:t-1}^3 \bigr)
\end{align}
\end{subequations}
To see why, consider for example the information known by node~1 at time $t$. Given $( x_{0:t}^1,\, x_{0:t-1}^2)$, we may use~\eqref{ex1_constraints} to compute past decisions $( u_{0:t-1}^1, u_{0:t-1}^2 )$. Then, using~\eqref{ex1_sseqn} we may infer the past disturbances $( w_{-1:t-1}^1,\, w_{-1:t-2}^2)$. Conversely, if $( w_{-1:t-1}^1,\, w_{-1:t-2}^2)$ is known, we may compute $( u_{0:t-1}^1, u_{0:t-1}^2 )$ via~\eqref{ex1_noiseform} and then compute $( x_{0:t}^1,\, x_{0:t-1}^2)$ via~\eqref{ex1_constraints}. It is straightforward to show that linearity of $\gamma$ implies linearity of $\hat\gamma$.

\subsection{State and input decomposition}\label{sec:stindecomp}

\begin{figure}[t]
\centering
\begin{tikzpicture}[>=latex]
\pgfdeclarelayer{background}
\pgfsetlayers{background,main}
\def\mdots{\,\,\cdot\cdot\cdot}
\def\dy{0.22}
\def\dax{1}
\def\dt{0.07}
\def\fs{\scriptsize}  
\tikzstyle{boxes}=[draw,fill=black!6,rectangle,thick,rounded corners=1mm,inner
sep=0ex]
\tikzstyle{labels}=[draw,fill=white,rectangle,thick,inner xsep=0.3mm,inner
ysep=0.7mm,anchor=north west]
\matrix (m) [matrix of math nodes, row sep=0.9em, column sep=0.9em, ampersand
replacement=\&]{
\text{Node $1$}	\& \mdots \& w^1_{t-3} \& w^1_{t-2} \& w^1_{t-1} \&  \\
\text{Node $2$}	\& \mdots \& w^2_{t-3} \& w^2_{t-2} \& w^2_{t-1} \&  \\
\text{Node $3$}	\& \mdots \& w^3_{t-3} \& w^3_{t-2} \& w^3_{t-1} \& \phantom{w^3_t} \\[1mm]
\& \& \text{\small $t-3$} \& \text{\small $t-2$} \& \text{\small $t-1$}\\ };
\draw [thick,->] (m-3-3.north west)+(-0.2,-\dax) --
	node[pos=0.8,anchor=north west,inner ysep=5pt] {Time} +(5,-\dax);
\draw [thick,dashedx] (m-3-3.north west)+(-0.2,-\dax) -- +(-0.9,-\dax);
\draw [thick] (m-3-3.north)+(0,-\dax-\dt) -- +(0,-\dax+\dt);
\draw [thick] (m-3-4.north)+(0,-\dax-\dt) -- +(0,-\dax+\dt);
\draw [thick] (m-3-5.north)+(0,-\dax-\dt) -- +(0,-\dax+\dt);
\draw [thick] (m-3-6.north)+(0,-\dax-\dt) -- +(0,-\dax+\dt);
\begin{pgfonlayer}{background}
	\node [boxes, fit=(m-1-5)]{};
	\node [boxes, fit=(m-2-5)]{};
	\node [boxes, fit=(m-1-4) (m-2-4) (m-2-2)] (AL) {};
	\node [boxes, fit=(m-3-5) (m-3-2)] (BL) {};
\end{pgfonlayer}
\path (m-1-5.north east)+(-\dy,\dy) node [labels] {\fs$\{1\}$};
\path (AL.north west)+(-\dy,\dy) node [labels] {\fs$\{1,2,3\}$};
\path (m-2-5.north east)+(-\dy,\dy) node [labels] {\fs$\{2,3\}$};
\path (BL.north west)+(-\dy,\dy) node [labels] {\fs$\{3\}$};
\end{tikzpicture}
\caption{Noise partition diagram for Example~\ref{ex:threeNode} (see
Fig.~\ref{fig:ex1_diagram}). The entire disturbance history is partitioned according to which subset of the nodes have access to the information. The subsets are indicated in the labels.
\label{fig:ex1_noisesplit}}
\if\MODE1\else\vspace{-3mm}\fi
\end{figure}
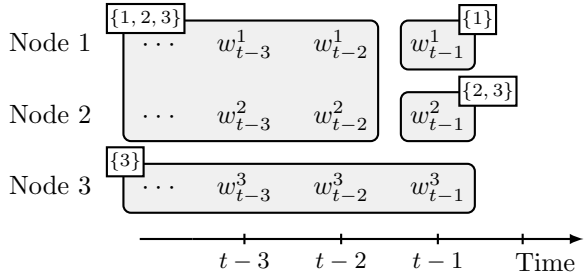

Extending the method for delays from \cite{lamperskidynamic2012}, we
regroup the disturbance terms in order to decompose
the input and state into independent random variables.
Note that \eqref{ex1_noiseform} can be used to partition the entire
noise history based on which subsets of the controllers can measure. This leads to the \emph{noise partition diagram}, as shown in Fig.~\ref{fig:ex1_noisesplit}. For example, the bottom cluster $\{\dots,w_{t-3}^3,w_{t-2}^3,w_{t-1}^3\}$ is available only to $u_t^3$, whereas the cluster $\{w_{t-1}^2\}$ is available to both $u_t^2$ and $u_t^3$. We call the noise subsets \emph{label sets} and denote them by $\Lab_t^s$, where
$s \in \{ \{1\},\{2,3\},\{3\},\{1,2,3\}\}$.
For example, $\Lab_t^{\{1\}} = \{ w_{t-1}^1 \}$. We may rewrite~\eqref{ex1_noiseform} as
\begin{subequations}\label{ex1_noiseform2}
\begin{align}
u^1_t &= \hat\gamma_t^1\bigl(  \Lab_t^{\{1\}}, \Lab_t^{\{1,2,3\}}  \bigr) \\
u^2_t &= \hat\gamma_t^2\bigl( \Lab_t^{\{2,3\}}, \Lab_t^{\{1,2,3\}} \bigr) \\
u^3_t &= \hat\gamma_t^3\bigl( \Lab_t^{\{3\}},\Lab_t^{\{2,3\}},\Lab_t^{\{1,2,3\}} \bigr)
\end{align}
\end{subequations}
Note that $u_t^i$ depends on $\Lab_t^s$ if and only if $i \in s$. Because the disturbances are mutually independent and the label sets are disjoint, we may decompose $u_t$ as a sum of its projections onto each of the $\Lab_t^s$. This leads to a decomposition of the form
\begin{equation}\label{ex1_inputdecomp}
u_t = \bmat{\varphi_t^{\{1\}} \\ 0 \\ 0}
+ \bmat{0\\ \bigl[\varphi_t^{\{2,3\}}\bigr]^2 \\ \bigl[\varphi_t^{\{2,3\}}\bigr]^3}
+ \bmat{0 \\ 0 \\ \varphi_t^{\{3\}}} + \varphi_t^{\{1,2,3\}}
\end{equation}
where $\varphi_t^s$ is a linear function of the elements of
$\Lab_t^s$. Note that under this decomposition, the $\varphi_t^s$
component of $u_t^i$ is zero if $i\notin s$. We shall see that the
states $x_t^i$ also depend linearly on the label sets in a manner
analogous to~\eqref{ex1_noiseform2}. Therefore, the state $x_t$ can be
similarly decomposed as
\begin{equation}\label{ex1_statedecomp}
x_t = \bmat{\zeta_t^{\{1\}} \\ 0 \\ 0}
+ \bmat{0\\ \bigl[\zeta_t^{\{2,3\}}\bigr]^2 \\ \bigl[\zeta_t^{\{2,3\}}\bigr]^3}
+ \bmat{0 \\ 0 \\ \zeta_t^{\{3\}}} + \zeta_t^{\{1,2,3\}}
\end{equation}
It will also be shown that the optimal decisions have the form $\varphi_t^s =
K_t^s \zeta_t^s$ where the $\{K_t^s\}$ are real matrices and the equivalent constraints from \eqref{ex1_constraints}, \eqref{ex1_noiseform}, and
\eqref{ex1_noiseform2} are satisfied by construction. 

\subsection{Update equations}

\begin{figure}[t]
\centering
\begin{tikzpicture}[thick,>=latex]
  \tikzstyle{loopstyle}=[out=90+35,in=90-35,loop,looseness=6]
  \tikzstyle{block}=[rectangle,draw]
  \def \n {3}
  \def \hgap {2cm}
  \def \vgap {30pt}
  \foreach \i in {1,...,\n}
  {
    \node (w\i) at ({(\i-1)*\hgap},0) {$w^{\i}$};
  }
  \node[block] (s1) at (0,\vgap) {$\{1\}$};
  \node[block] (s23) at (\hgap,\vgap) {$\{2,3\}$};
  \node[block] (s3) at (2*\hgap,\vgap) {$\{3\}$};

  \node[block] (s123) at ({\hgap/2},2*\vgap) {$\{1,2,3\}$};

  \draw[->,dashedx] (w1) edge (s1);
  \draw[->,dashedx] (w2) edge (s23);
  \draw[->,dashedx] (w3) edge (s3);

  \draw[->] (s1) edge  (s123);
  \draw[->] (s23) edge (s123);

  \draw[->] (s123) edge [loopstyle] (s123);
  \draw[->] (s3) edge [loopstyle] (s3);

\end{tikzpicture}
\caption{\label{fig:ex1_hierarchy} Information graph for Example \ref{ex:threeNode}. Each node is a label set, which is a subsets of nodes in the network graph (see Fig.~\ref{fig:ex1_diagram}).}
\if\MODE1\else\vspace{-3mm}\fi
\end{figure}
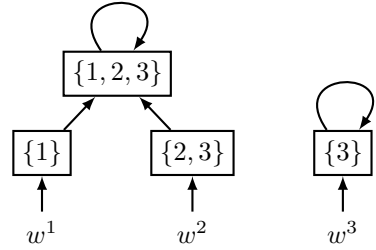

The optimality proof for our controller uses dynamic
programming and requires a description of the evolution of $\zeta_t^s$ over time. Since $\zeta_t^s$
and $\phi_t^s$ are linear functions of the label set $\Lab_t^s$ terms,
the dynamics of the label sets will be described as an intermediate
step.
From the noise partition diagram of Fig.~\ref{fig:ex1_noisesplit}, the
label sets have dynamics
\begin{equation}\label{ex1_labeldynamics}
\begin{aligned}
\Lab_{t+1}^{\{1\}}&= \{w_t^1\}, & \Lab_{t+1}^{\{3\}} &=   \Lab_t^{\{3\}} \cup\{w_1^3\},\\
\Lab_{t+1}^{\{2,3\}} &= \{w_t^2\}, & \Lab_{t+1}^{\{1,2,3\}} &= \Lab_t^{\{1,2,3\}} \cup \Lab_t^{\{1\}} \cup
\Lab_t^{\{2,3\}}
\end{aligned}
\end{equation}
with initial conditions
\begin{align*}
\Lab_0^{\{1\}} &= \{x_0^1\}, &
\Lab_0^{\{3\}} &= \{x_0^3\}, \\
\Lab_0^{\{2,3\}} &= \{x_0^2\}, &
\Lab_0^{\{1,2,3\}} &= \emptyset.
\end{align*}
A convenient way of visualizing the label set
dynamics~\eqref{ex1_labeldynamics} is by using an \emph{information
  graph} as shown in Fig.~\ref{fig:ex1_hierarchy} (cf. \cite{lamperskidynamic2012}). An edge $r \to s$ indicates that $\Lab_t^r \subset \Lab_{t+1}^s$. Similarly, an edge $w^i \to s$ indicates that $\{w_t^i\} \subset \Lab_{t+1}^s$.
The information graph also illustrates the propagation of disturbances via~\eqref{ex1_sseqn}. For example, $w^2$ is injected at node~2, affects nodes $\{2,3\}$ immediately, and affects $\{1,2,3\}$ after one timestep and for every timestep thereafter.
It can be shown by induction that the $\zeta_t$ coordinates defined below
satisfy \eqref{ex1_statedecomp} for $t=0,\ldots,T$.
\begin{subequations} \label{ex1_zetadynamics}
\begin{align} 
\zeta_{t+1}^{\{1\}} &= w_t^1 \label{q1}\\
\zeta_{t+1}^{\{2,3\}} &= \bmat{w_t^2\\ 0} \label{q2}\\
\zeta_{t+1}^{\{3\}} &= A_t^{33} \zeta_t^{\{3\}} +B_t^{33}
\varphi_t^{\{3\}} + w_t^3 \label{q3}\\
\zeta_{t+1}^{\{1,2,3\}} &= 
A_t \zeta_t^{\{1,2,3\}}+B_t \phi_t^{\{1,2,3\}} + \eta_t \label{q4}
\end{align}
\end{subequations}
with initial conditions
\begin{equation*}
\zeta_0^{\{1\}} = x_0^1,\:\: 
\zeta_0^{\{2,3\}} = \bmat{x_0^2 \\ 0}, \:\:
\zeta_0^{\{3\}} = x_0^3, \:\:
\zeta_0^{\{1,2,3\}} = 0
\end{equation*}
and where we have defined
\begin{multline*}
\eta_t \defeq A_t^{\{1,2,3\},\{2,3\}} \zeta_t^{\{2,3\}}
 +B_t^{\{1,2,3\},\{2,3\}} \varphi_t^{\{2,3\}}\\
 + A_t^{\{1,2,3\},\{1\}} \zeta_t^{\{1\}}
 +B_t^{\{1,2,3\},\{1\}} \varphi_t^{\{1\}}.
\end{multline*}
The notation $A_t^{rs}$ denotes the submatrix $[A_t^{ij}]$ with $i\in r$ and $j\in s$. For example,
$
A_t^{\{3\},\{2,3\}} = \bmat{A_t^{32} & A_t^{33}}
$.

Note that the dynamics in \eqref{ex1_zetadynamics} can be deduced directly from the information graph. Indeed, $\zeta_{t+1}^s$ only
depends on $w_t^i$ if $w^i\to s$ and $\zeta_{t+1}^s$ only depends on
$(\zeta_t^r,\varphi_t^r)$ whenever $r\to s$. This leads to the following compact
representation of the dynamics.
\begin{subequations}\label{eq:zeta_both}
\begin{align} 
\zeta_0^s &= \sum_{w^i \to s} I^{s,\{i\}} x_0^i \label{eq:zetaInit}\\ \label{eq:zeta}
\zeta_{t+1}^s &= \sum_{ r \to s} \bigl(A_t^{s r}\zeta_t^r+B_t^{s r}\phi_t^r\bigr)+ \sum_{w^i \to s}I^{s,\{i\}}w_t^i .
\end{align}
\end{subequations}

\subsection{Decoupled Optimization Problems}

Using the theory developed so far, we will sketch the strategy for
decoupling optimization problems. The method
is based on dynamic programming. 

Suppose that the expected cost incurred by the optimal policy $\gamma_{0:T-1}^*$ for steps $t+1,\ldots, T$ has the
form
\begin{multline}\label{costDecoupleSketch}
\ee^{\gamma^*}\Biggl(
\sum_{k=t+1}^{T-1} \bmat{x_k\\u_k}^\tp \bmat{Q_k & S_k \\ S_k^\tp & R_k} \bmat{x_k\\u_k}
+ x_T^\tp Q_f x_T^{\vphantom{\tp}}
\Biggr)
\\
= \sum_{s} \ee^{\gamma^*}\Bigl(
\left(\zeta_{t+1}^s\right)^\tp X_{t+1}^s \zeta_{t+1}^s
\Bigr)+c_{t+1}
\end{multline}
where the $X_{t+1}^s$ are positive semidefinite, $c_{t+1}$ is a constant, and the sum ranges over the nodes of the information graph from
Fig.~\ref{fig:ex1_hierarchy}. Using \eqref{ex1_statedecomp}, this
decomposition holds at $t+1=T$ with $X_T^s=Q_f^{ss}$ and $c_{T} = 0$. 

Substituting \eqref{eq:zeta_both} into~\eqref{costDecoupleSketch} and using
independence, the expected cost for steps $t,t+1,\ldots,T$ is given by
\begin{equation}\label{costToGoSketch}
\sum_{r} \ee^{\gamma^*}\Biggl(
\bmat{\zeta_t^r\\\phi_t^r}^\tp
\Gamma_t^r
\bmat{\zeta_t^r\\ \phi_t^r}
\Biggr)+c_t,
\end{equation}
where $r$ ranges over all nodes in the information graph and
$\Gamma_t^r$ and $c_t$ are given by
\begin{align}
\Gamma_t^r  &= \bmat{Q_t^{rr} & S_t^{rr} \\ {S_t^{rr}}^\tp & R_t^{rr}} +
\bmat{A_t^{sr} & B_t^{sr}}^\tp X_{t+1}^s \bmat{A_t^{sr} & B_t^{sr}} \\
c_t &= c_{t+1} + \sum_{\substack{i \in \mathcal{V}\\ w^i \to s}}
\Trace\left(  (X_{t+1}^s)^{\{i\},\{i\}} W_t^i\right). \label{eq:cost_recursion}
\end{align}
Here $s$ is the unique node such that $r\to s$. 

Note that $\Gamma_t^r$ is positive semidefinite, with a positive
definite lower right block. It follows that the quadratic form in
\eqref{costToGoSketch} is minimized over $\phi_t^r$ by a linear
mapping 
\begin{equation}\label{gainSketch}
\phi_t^r = K_t^r \zeta_t^r. 
\end{equation}
As discussed in Section~\ref{sec:stindecomp}, the mapping~\eqref{gainSketch}
satisfies the information constraints of the problem. Furthermore, the optimal cost for time steps $t,t+1,\ldots,T$ is of the form \eqref{costDecoupleSketch}. 

\subsection{Message passing implementation}

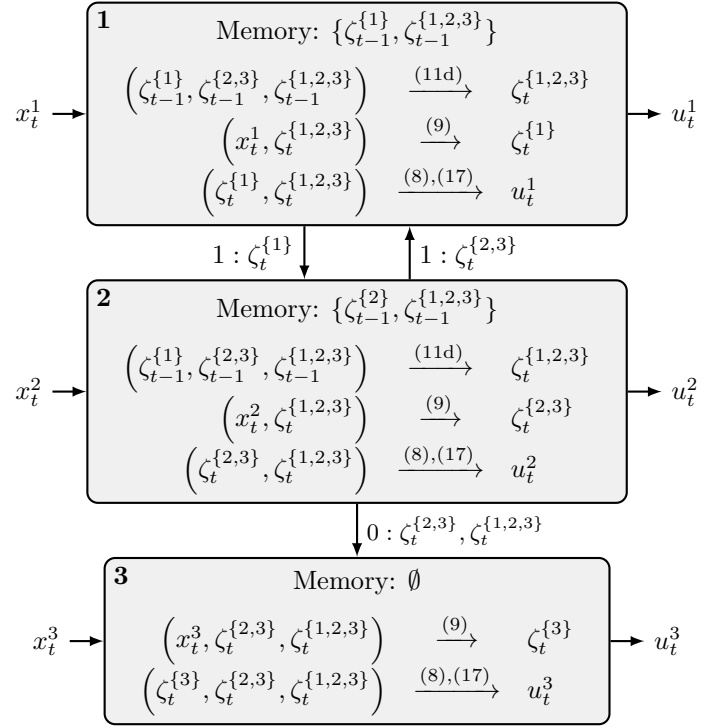
\begin{figure}[ht]
\centering
\begin{tikzpicture}[thick,node distance=3cm,>=latex]
\tikzstyle{block}=[rectangle,draw,minimum width=64mm, fill=black!6, rounded corners]
\def\fs{\small}
\def\h{4.5mm}		
\def\del{22mm}		
\def\w{7mm}			

\node[block,anchor=south] (Sys1) at (0,\del) {%
$
\begin{array}{c}
\textrm{Memory: } \{\zeta_{t-1}^{\{1\}},\zeta_{t-1}^{\{1,2,3\}}\}
\vspace{2mm}
\\
\begin{array}{rcl}
\left(
   \zeta_{t-1}^{\{1\}},\zeta_{t-1}^{\{2,3\}},\zeta_{t-1}^{\{1,2,3\}}
   \right) &\xrightarrow{\eqref{q4}} &\zeta_t^{\{1,2,3\}} 
\\
\left( x_t^1, \zeta_t^{\{1,2,3\}} \right) &
\xrightarrow{\eqref{ex1_statedecomp}} &\zeta_t^{\{1\}} \\
\left(\zeta_t^{\{1\}}, \zeta_t^{\{1,2,3\}} \right) & 
\xrightarrow{\eqref{ex1_inputdecomp},\eqref{gainSketch}} & u_t^1
\end{array}
\end{array}
$};
\node[block] (Sys2) at (0,0) {%
$
\begin{array}{c}
\textrm{Memory: } \{\zeta_{t-1}^{\{2\}},\zeta_{t-1}^{\{1,2,3\}} \}
\vspace{2mm}
\\ 
\begin{array}{rcl}
\left(\zeta_{t-1}^{\{1\}}, \zeta_{t-1}^{\{2,3\}},
  \zeta_{t-1}^{\{1,2,3\}}\right) 
&\xrightarrow{\eqref{q4}} & \zeta_t^{\{1,2,3\}} \\
\left(x_t^2, \zeta_t^{\{1,2,3\}}\right) 
&\xrightarrow{\eqref{ex1_statedecomp}} & \zeta_t^{\{2,3\}} \\
\left(\zeta_t^{\{2,3\}},\zeta_t^{\{1,2,3\}}\right) 
&\xrightarrow{\eqref{ex1_inputdecomp},\eqref{gainSketch}} & u_t^2
\end{array}
\end{array}
$};
\node[block,anchor=north] (Sys3) at (0,-\del) {%
$
\begin{array}{c}
\textrm{Memory: } \emptyset
\vspace{2mm}
\\
\begin{array}{rcl}
\left(x_t^3,\zeta_t^{\{2,3\}},\zeta_t^{\{1,2,3\}} \right) 
&\xrightarrow{\eqref{ex1_statedecomp}} & \zeta_t^{\{3\}} \\
\left(\zeta_t^{\{3\}},\zeta_t^{\{2,3\}},\zeta_t^{\{1,2,3\}} \right) 
&\xrightarrow{\eqref{ex1_inputdecomp},\eqref{gainSketch}} & u_t^3 
\end{array}
\end{array}
$};
\draw[<-] (Sys1.west) -- +(-\h,0) node[anchor=east] {$x_t^1$};
\draw[<-] (Sys2.west) -- +(-\h,0) node[anchor=east] {$x_t^2$};
\draw[<-] (Sys3.west) -- +(-\h,0) node[anchor=east] {$x_t^3$};
\draw[->] (Sys1.east) -- +(\h,0) node[anchor=west] {$u_t^1$};
\draw[->] (Sys2.east) -- +(\h,0) node[anchor=west] {$u_t^2$};
\draw[->] (Sys3.east) -- +(\h,0) node[anchor=west] {$u_t^3$};

\draw[->] (Sys2) -- node[right]{\fs$0:\zeta^{\{2,3\}}_t,\zeta^{\{1,2,3\}}_t$} (Sys3);
\draw[->,transform canvas={xshift=-\w}] (Sys1) -- node[left] {$1:\zeta_t^{\{1\}}$} (Sys2);
\draw[<-,transform canvas={xshift=\w}] (Sys1) -- node[right] {$1:\zeta_t^{\{2,3\}}$} (Sys2);

\node[anchor=north west] at (Sys1.north west) {\bf 1};
\node[anchor=north west] at (Sys2.north west) {\bf 2};
\node[anchor=north west] at (Sys3.north west) {\bf 3};
\node (spacer) at (Sys3.south) {\vphantom{x}};

\end{tikzpicture}
\caption{Network graph for Example~\ref{ex:threeNode} with messages
  and memory. The syntax $(\dots) \to \zeta_t^s$ means that
  $\zeta_t^s$ is computed  as a function of the left-hand side terms. 
  The numbers above the arrows indicate which
  equations are involved in the computation. 
The messages passed between nodes are shown next to the graph edges.
\label{fig:ex1_messages}}
\end{figure}

Implementing the controller described above is not straightforward
because state information must be used to deduce past
disturbances. Using a combination of local measurements, local memory,
and message passing, the optimal inputs can be computed in a
distributed fashion without computing past noise terms.
The proposed implementation may be visualized by augmenting Fig.~\ref{fig:ex1_diagram} to include the appropriate messages, memory, and update equations. See Fig.~\ref{fig:ex1_messages}.

\if\MODE1\bigskip\fi

\if\MODE1\vspace{2mm}\else\fi

\section{Problem statement: the general case}\label{sec:prob_statement}

We begin by defining some useful notation.  
The symbol $I$ denotes a block-identity matrix whose dimensions are to be inferred by context. This notation is useful for extracting blocks from larger matrices. For example, if $A_t$ is as in Example~\eqref{ex:threeNode}, the fact that $A_t^{13}=0$ and $A_t^{23}=0$ implies that
\[
A_t I^{\{1,2,3\},\{3\}} = I^{\{1,2,3\},\{3\}} A_t^{33}
\]
If $\mathcal{Y} = \{y^1,\ldots,y^M\}$ is a set of random vectors (possibly of different sizes), we say that
$
z\in \Span\mathcal{Y}
$
if there are appropriately sized real matrices $C^1,\ldots, C^M$ such that 
$
z = C^1 y^1 + \dots + C^M y^M
$.

We also require some basic definitions regarding graphs. A
\emph{network graph} $G(\mathcal{V},\mathcal{E})$ is a directed graph
where each edge is labeled with a~$0$ if the associated link is
delay-free, or a~$1$ if it has a one-timestep delay. The vertices are
$\mathcal{V}=\{1,\dots,n\}$. If there is an edge from $j$ to $i$, we
write $(j,i)\in \mathcal{E}$, or simply $j\to i$. When delays are
pertinent, they are denoted as $j\nodelay i$ or $j\delay i$. Directed
cycles are permitted in the network graph, but we assume there are no
directed cycles with a total delay of zero. In our framework, all
nodes belonging to such a delay-free cycle can be collapsed into a
single node. Note
that Fig.~\ref{fig:ex1_diagram} shows the network graph for
Example~\ref{ex:threeNode}. 
Associated with the network graph $G(\mathcal{\mathcal{V}},\mathcal{E})$ is the \emph{delay matrix}~$D$. Each entry~$D^{ij}$ is the sum of the delays along the directed path from $j$ to $i$ with the shortest aggregate delay. We assume~$D^{ii}=0$ for all~$i$, and if no directed path exists, we set $D^{ij}=\infty$. The delay matrix for Example~\ref{ex:threeNode} is
\begin{equation}\label{ex1_delaymatrix}
D = \bmat{0 & 1 & \infty \\ 1 & 0 & \infty \\ 1 & 0 & 0}.
\end{equation}
Delays are assumed to be fixed for all time. 

We now state the general class of problems that can be solved using the method developed in this paper.

\begin{prob}\label{probmain}
Let $G(\mathcal{V},\mathcal{E})$ be a network graph with associated delay matrix $D$. Suppose the following time-varying state-space equations are given for all $i\in\mathcal{V}$ and for $t=0,\dots,T-1$.
\begin{equation}\label{eq:ss_eqns}
x_{t+1}^i = \sum_{\substack{j\in \mathcal{V}\\D^{ij}\leq 1}}
\bigl( A^{ij}_t x_t^j + B^{ij}_t u_t^j \bigr) + w_t^i
\end{equation}
Stacking the various vectors and matrices, we obtain the more compact representation
\begin{equation}\label{eq:ss_simplest}
x_{t+1} = A_tx_t + B_tu_t + w_t.
\end{equation}
All random disturbances are assumed to be jointly Gaussian and independent from one another. Specifically, the random vectors
$
\{ x_0^i, w_0^i, \dots, w_{T-1}^i \}_{i\in \mathcal{V}}
$
are mutually independent. Their means and covariances are given by~\eqref{ass:norm_distributions}.
The inputs~$u_{0:T-1}$ are controlled using state feedback subject to an information constraint. The \emph{information set} for controller~$i$ at time~$t$ is as follows.
\begin{equation}\label{infoset}
\Info_t^i = \bigl\{x_k^j:j\in \mathcal{V},\:\: 0\le k \le t-D^{ij}\bigr\}.
\end{equation}
So $\Info_t^i$ is the set of states belonging to nodes that have had sufficient time to reach node~$i$ by time~$t$. Each decision-maker measures its corresponding information set
\begin{equation}
\label{eq:info}
u_t^i = \gamma_t^i( \Info_t^i ).
\end{equation}
The goal is to choose the set of policies $\gamma = \{\gamma_{0:T-1}^i\}_{i\in \mathcal{V}}$ that minimize the expected quadratic cost~\eqref{eq:cost_function}
\end{prob}
In Problem~\ref{probmain}, as in Example~\ref{ex:threeNode}, we assume
that all decision-makers know the system dynamics, cost matrices, and
network graph. 

The problem is cooperative in nature; we are to jointly design the set
of policies~$\gamma$ to optimize the
cost~\eqref{eq:cost_function}. Based on our formulation, the
controller may be any function of the past information history, which
grows with the size of the time horizon~$T$. We will show by
construction that there exists an optimal policy that has a finite
memory that is independent of~$T$.

\begin{figure}[bht]
\centering
\begin{tikzpicture}[thick,node distance=1.7cm,>=latex]
	\def\fs{\small}  
	\tikzstyle{block}=[circle,draw,fill=black!6,minimum height=2.1em]
	\tikzstyle{loopstyle}=[loop,looseness=8]
	\node [block](N1){$1$};
	\node [block, right of=N1](N2){$2$};
	\draw [->] (N1) edge node[above]{\fs$2$} (N2);
\end{tikzpicture}
\qquad\quad
\begin{tikzpicture}[thick,node distance=1.7cm,>=latex]
	\def\fs{\small}  
	\tikzstyle{block}=[circle,draw,fill=black!6,minimum height=2.1em]
	\tikzstyle{loopstyle}=[loop,looseness=8]
	\node [block](N1){$1$};
	\node [block, right of=N1](Nr){$R$};
	\node [block, right of=Nr](N2){$2$};
	\draw [->] (N1) edge node[above]{\fs$1$} (Nr);
	\draw [->] (Nr) edge node[above]{\fs$1$} (N2);
\end{tikzpicture}
\caption{Network graph for a two-timestep delay. The equivalent representation on the right uses a relay node and two one-timestep delays.
\label{fig:relay}}
\end{figure}
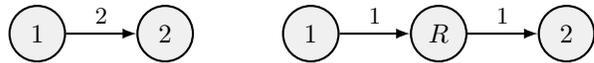
\if\MODE1\textbf{Larger delays. }\else\paragraph{Larger delays.}\fi
The problem formulation considered in this paper only allows for delays of $0$ or $1$ timestep along edges of the network graph. Larger delays can be accommodated by including \emph{relay} nodes. Consider for example the network graph of Fig.~\ref{fig:relay}.
The relay node has trivial dynamics. At time $t$, the relay node
stores $x_{t-1}^1$. Therefore, at time $t$, node 2 receives
$x_{t-2}^1$ from the relay node, as desired. The system matrices for
the relay node are also chosen so that there is no injected noise and
no cost incurred. While any fixed delay can be simulated with an
appropriate number of relay nodes, large delays will
result in high dimensional controllers. 
Alternative methods to handle
large delays is an area of future work.

\if\MODE1\else\vspace{2mm}\fi
\section{Main results}\label{sec:main}

This section presents the main result: an explicit state-space
solution for Problem~\ref{probmain}. Two equivalent forms of the controller will be presented.
The first form has states which are functions of the primitive random variables $x_0$ and $w_{0:t-1}$. The second form uses state feedback only and gives a distributed implementation that uses message passing. 

Our first step is to define the information graph (as in
Fig.~\ref{fig:ex1_hierarchy}) for the general graph case (cf. \cite{lamperskidynamic2012}). Let $s_k^j$ be the set of nodes reachable from node $j$ within $k$ steps: 
\begin{equation}\label{eq:sjk}
s_k^j = \{i\in \mathcal{V}: D^{ij} \le k\}.
\end{equation}
The information graph $\hat G(\mathcal{U},\mathcal{F})$, is given by 
\begin{align*}
\mathcal{U} &= \{s_k^j:k\ge 0,\:\: j\in \mathcal{V}\} \\
\mathcal{F} &= \{(s_{k}^j,s_{k+1}^j):k\ge 0,\:\: j\in \mathcal{V}\}. 
\end{align*}
The additional labels~$w^i$ are not counted amongst the nodes of $\hat
G$ as a matter of convention, but are shown as a reminder of which
noise signal is being tracked. We will often write expressions such as
$\{s\in\mathcal{U}: w^i\to s\}$ to denote the set of root nodes
of~$\hat G$. The following proposition gives some useful properties of
the information graph.

\begin{prop}\label{prop:hierarchy_properties}
Given an information graph $\hat G(\mathcal{U},\mathcal{F})$, the following properties hold.
\begin{enumerate}[(i)]
\itemsep=1mm
\item Every node in $\hat G$ has exactly one descendant. In other words, for every $ r \in\mathcal{U}$, there is a unique $s \in\mathcal{U}$ such that $ r \to s$.
\item Every path eventually hits a node with a self-loop.
\item If the network graph satisfies $|\mathcal{V}|=n$, the number of nodes in $\hat G$ is bounded by
$
n \le |\mathcal{U}| \le n^2-n+1.
$
\end{enumerate}
\end{prop}

The first two properties are immediate by construction. The lower bound
on~$|\mathcal{U}|$ is achieved by directed acyclic network graphs with zero
delay on all edges. The upper bound on~$|\mathcal{U}|$ is achieved by
network graph consisting of one large cycle: $1\to 2 \to \dots \to n \to
1$, and each link has a one-timestep delay. 
Note that the information graph may have several connected
components. This happens whenever the network graph is not strongly
connected. For example, Fig.~\ref{fig:ex1_hierarchy} has two
connected components because there is no path $3\to 2$ in
Fig.~\ref{fig:ex1_diagram}. 

We are now ready to present the main result of this paper, which
expresses the optimal controller as a function of new coordinates
induced by the information graph. 

\begin{thm}\label{thm:main}
Consider Problem~\ref{probmain}, and let $\hat G(\mathcal{U},\mathcal{F})$ be the associated information graph. Define the matrices $\{X^r_{0:T}\}_{r\in\mathcal{U}}$ and $\{K^r_{0:T-1}\}_{r\in\mathcal{U}}$ recursively as follows,
\begin{subequations}\label{eq:X-recursion}
\begin{align}
X_T^ r &= Q_f^{ r r} \\
\Omega_t^r &= R^{rr}_t+{B^{sr}_t}^{\tp}X_{t+1}^sB^{sr}_t \\
K_t^r &= -\bigl(\Omega_t^{r}\bigr)^{-1}
\bigl(S_t^{ r r}+{A_t^{s r}}^\tp X_{t+1}^s B_t^{s r}\bigr)^\tp \\
\label{eq:X-Riccati}
X_t^r &= Q_t^{ r r}+{A_t^{s r}}^\tp X_{t+1}^s A_t^{s r}
-{K_{t}^r}^{\tp} \Omega_t^r K_t^r
\end{align}
\end{subequations}
where for each $r\in\mathcal{U}$, we have defined $s \in\mathcal{U}$ to be the unique node such that $r \to s$. The optimal control decisions satisfy the following state-space equations
\begin{subequations}\label{eq:ss_sol}
\begin{align} 
\zeta_0^s &= \sum_{w^i \to s} I^{s,\{i\}} x_0^i \\
\label{eq:ss_update}
\zeta_{t+1}^s &= \sum_{ r \to s} \bigl(A_t^{s r}+B_t^{s r}K_t^r\bigr) \zeta_t^r + 
\sum_{w^i \to s}I^{s,\{i\}}w_t^i \\
u_t^i &= \sum_{r\ni i} I^{\{i\},r} K_t^r \zeta_t^r.
\end{align}
\end{subequations}
The corresponding optimal expected cost is
\begin{multline}\label{eq:opt_cost}
V_0 = \sum_{\substack{i\in \mathcal{V}\\w^i\to s}}
\Trace\left((X_{0}^s)^{\{i\},\{i\}} \Sigma_0^s \right) \\[-2mm]
+ \sum_{t=0}^{T-1}\sum_{\substack{i\in \mathcal{V}\\w^i\to s}} \Trace\Bigl(
(X_{t+1}^s)^{\{i\},\{i\}}W_t^i \Bigr).
\end{multline}\vspace{-7mm}
\end{thm}
\begin{pf}
See Section~\ref{sec:mainproof}.
\end{pf}
Note that when $r\in\mathcal{U}$ has a self-loop, the recursion for
$X_t^r$ only depends on $X_{t+1}^r$ and is a classical Riccati
equation. 
Otherwise, repeated application of \eqref{eq:X-recursion} shows that
$X_{t}^r$ is a function of $X_{t+k}^s$, where $s\to s$ is
the unique self loop reachable from $r$ and $k$ is the length of the path. 

Equation~\eqref{eq:ss_sol} expresses the controller as a map
$w\mapsto u$. Our second main result, Theorem~\ref{thm:msg}, gives a message passing
implementation of the optimal controller from Theorem~\ref{thm:main}
that explicitly finds the map $x\mapsto u$.

\begin{thm}\label{thm:msg}
Consider the problem setting of Theorem~\ref{thm:main}.
For each node $i$ and all $t=0,\ldots,T-1$, define the outgoing
message sent from node~$i$ to node~$j$ by
\if\MODE1\vspace{-2mm}\fi
\begin{subequations}\label{eq:msgDef}
\begin{align}
\text{If }i\nodelay j:\quad \Msg_t^{ij} &=  \{\zeta_t^s: s\in
\mathcal{U}, \, i,j\in s\}. \\
\text{If }i\delay j:\quad \Msg_t^{ij} &=  \{\zeta_t^s: 
s\in \mathcal{U}, \,
i\in
s,\:\: j\notin s\}.
\end{align}
\end{subequations}
and define the local memory of node $i$ by 
\begin{subequations}\label{eq:memDef}
\begin{align}
\Mem_0^i &= \emptyset \\
\Mem_{t+1}^i &= \{\zeta_t^s:s\in\mathcal{U},\, i \in s,\, \nexists
               j\in s \textrm{ with } j\nodelay i \}
\end{align}
\end{subequations}
If controller $i$ measures $x_t^i$ at time $t$, then the distributed algorithm
defined by \eqref{eq:msgDef} and \eqref{eq:memDef} can be executed
without deadlock. In other words, the $\mathcal{M}_t^{ij}$ and $\mathcal{R}_t^i$ can be
computed for all $t$ and $i$.  Furthermore, if $i\in s\in \mathcal{U}$ then 
\begin{equation}\label{messagePassingSpan}
\zeta_t^s \in \Span\Biggl(
\{x_t^i\}\cup \Mem_t^i \cup \bigcup_{j\nodelay i} \Msg_t^{ji} 
\cup \bigcup_{j\delay i} \Msg_{t-1}^{ji}
\Biggr),
\end{equation}
where $\Msg_{-1}^{ji} = \emptyset$. Thus, the optimal $u_t^i$ at every timestep can be computed from the measurement, the local memory, and the incoming messages at time $t$.
\end{thm}
\if\MODE1\vspace{-3mm}\fi
\begin{pf}
See Section~\ref{pf:msg}.
\end{pf}
We will prove in Section~\ref{sec:mainproof} that the optimal
controller is unique. However, the choice of realization is not
unique, and there is no guarantee that the representation given in Theorem~\ref{thm:msg} will be minimal.

It is expected that constraining the information available to each node should result in an increase in computational and storage burden. Indeed, the memory required by each node in Theorem~\ref{thm:msg} may be large because it depends on how many $s\in\mathcal{U}$ contain~$i$. If the global state $x_t$ has dimension $N$ and there are $n$ nodes, the memory is bounded by $|\Mem_t^i| \le n^2N$. Note that this bound is independent of the horizon length $T$.

\subsection{Extension to the infinite-horizon case}\label{sec:infinte_horizon}

Our solution extends naturally to an infinite horizon, as well. To this end, we assume all system parameters are time-invariant. That is,  $A_0=A_1=\dots=A$ and similarly for $B$, $Q$, $R$, $S$, $W$. We seek a stabilizing controller that minimizes the average finite-horizon cost as the length of the horizon tends to infinity. Specifically, we replace the cost~\eqref{eq:cost_function} by
\begin{equation}\label{eq:cost_infinite}
\min_\gamma 
 \lim_{T\to\infty} \ee^\gamma \Biggl(
\frac{1}{T}\sum_{t=0}^{T-1} \bmat{x_t\\u_t}^\tp \bmat{Q & S \\ S^\tp & R} \bmat{x_t\\u_t}
\Biggr)
\end{equation}

\begin{cor}\label{cor:inf}
Consider Problem~\ref{probmain} under the time-invariance and average cost assumptions above and let $\hat G(\mathcal{U},\mathcal{F})$ be the associated information graph. Further suppose that for each self-loop $s\to s$ in the information graph, the following assumptions hold:
\begin{enumerate}[(1)]
\item $(A^{ss},B^{ss})$ is stabilizable
\item $\bmat{A^{ss} - e^{\jmath\theta}I & B^{ss}\\ C^{ss} & D^{ss} }$ has full column rank $\forall\theta \in [0,2\pi]$
\end{enumerate}
where $C^{ss}$ and $D^{ss}$ are any matrices that factorize
\[
\bmat{ Q^{ss} &  S^{ss} \\ {S^{ss}}^\tp &  R^{ss} } = 
\bmat{ C^{ss} &  D^{ss} }^\tp \bmat{ C^{ss} &  D^{ss} }
\]
Define the matrices $\{X^r\}_{r\in\mathcal{U}}$ and $\{K^r\}_{r\in\mathcal{U}}$ as follows
\begin{subequations}\label{eq:X-recursion_inf}
\begin{align}
\Omega^r &= R^{rr}+{B^{sr}}^{\tp}X^sB^{sr} \\
K^r &= -\bigl(\Omega^{r}\bigr)^{-1}
\bigl(S^{ r r}+{A^{s r}}^\tp X^s B^{s r}\bigr)^\tp \\
\label{eq:Inf-Riccati}
X^r &= Q^{ r r}+{A^{s r}}^\tp X^s A^{s r}
-{K^r}^{\tp} \Omega^r K^r
\end{align}
\end{subequations}
where for each $r\in\mathcal{U}$, we have defined $s \in\mathcal{U}$ to be the unique node such that $r \to s$.
The optimal steady-state controller satisfies the following state-space equations
\begin{subequations}\label{eq:ss_sol_inf}
\begin{align} 
\zeta_{t+1}^s &= \sum_{ r \to s} \bigl(A^{s r}+B^{s r}K^r\bigr) \zeta_t^r + 
\sum_{w^i \to s}I^{s,\{i\}}w_t^i \\
u_t^i &= \sum_{r\ni i} I^{\{i\},r} K^r \zeta_t^r
\end{align}
\end{subequations}
The corresponding optimal expected average cost is
\begin{equation}\label{eq:opt_cost_inf}
V_0 = \sum_{\substack{i\in \mathcal{V}\\w^i\to s}} \Trace\Bigl(
(X^s)^{\{i\},\{i\}}W^i \Bigr)
\end{equation}
\end{cor}

\begin{pf}
If $s\to s$ is a self-loop, then \eqref{eq:X-Riccati} reduces to a
classical Riccati difference equation, uncoupled from the other nodes
in the information graph. The conditions in the corollary guarantee
that for any fixed $t$, when $T\to \infty$, the value of $X_t^s$ converges to a
stabilizing solution to the corresponding algebraic Riccati equation
\eqref{eq:Inf-Riccati}, \cite{zdg}. In this case, $A^{ss}+B^{ss}K^s$ is stable. 

If $r$ is not a self-loop, then there is a unique node
$s\in\mathcal{U}$ such that the path from $r$ leads to $s$ and $s$ has
a self-loop. Since $R^{rr}$ is positive definite, \eqref{eq:X-Riccati}
implies that $X_t^r$ is a continuous function of $X_{t+k}^s$, where
$k$ is the length of the path from $r$ to $s$ in the information
graph. Thus $X_t^r$ approaches the constant value $X^r$, as
$X_t^s\to X^s$. Similarly, the optimal gains converge to $K^r$ as
$T\to\infty$. The corresponding cost follows by dividing the right of
\eqref{eq:opt_cost} by $T$ and taking the limit as $T\to\infty$. 

To see that the controller is stabilizing, it suffices to show that
\eqref{eq:ss_sol_inf} defines a stable mapping $w\mapsto \zeta$. Note
that if $s$ is not a self-loop, then $\zeta_t^s$ depends 
on only a finite number of noise terms, so the mapping $w\mapsto\zeta$
is stable. If $s$ is a self-loop, then the terms 
\begin{equation*}
\sum_{\substack{r\to s \\ r\ne s}} \bigl(A^{s r}+B^{s r}K^r\bigr) \zeta_t^r + 
\sum_{w^i \to s}I^{s,\{i\}}w_t^i
\end{equation*}
can be represented by filtering $w_t$ with a finite impulse response (FIR)
transfer function. Since $A^{ss}+B^{ss}K^{s}$ is stable, 
\eqref{eq:ss_sol_inf} reduces to a stable system which is forced by
FIR colored noise. Thus, the overall system is stable. 
\if\MODE1\qed\fi
\end{pf}

The message passing implementation of Theorem~\ref{thm:msg} can also be extended to the infinite-horizon. For a numerical example and further discussion, see Section~\ref{sec:example}. 

\section{Specialization to existing results}\label{sec:discussion1}

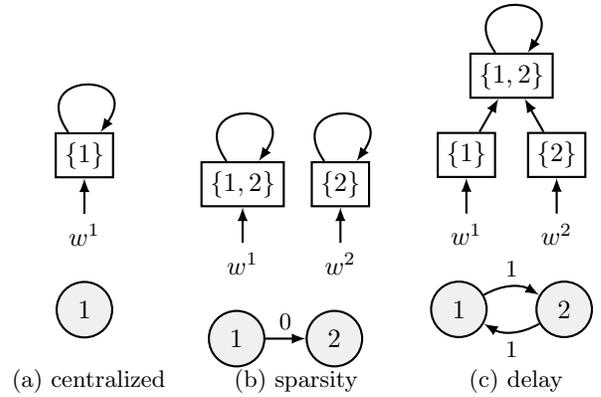
\begin{figure}[ht]
\centering 
\begin{subfigure}[b]{0.26\linewidth}
	\centering
	\begin{tikzpicture}[thick,node distance=2.00cm,>=latex]
	\begin{scope}
	  \def\fs{\small}  
	  \tikzstyle{block}=[circle,draw,fill=black!6,minimum height=2.1em]
	  \tikzstyle{loopstyle}=[loop,looseness=8]
	  \node [block] (N1){$1$};
	\end{scope}
	\begin{scope}[shift={(0,1)}]
	  \tikzstyle{loopstyle}=[out=90+35,in=90-35,loop,looseness=6]
	  \tikzstyle{block}=[rectangle,draw]
	  \def \vgap {30pt};
	  \def \hgap {1.0cm}
	  \node (w1) at (0,0) {$w^1$};
	  \node[block] (s1) at (0,\vgap) {$\{1\}$};
	  \draw[->,dashedx] (w1) edge (s1);
	  \draw[->] (s1) edge [loopstyle] (s1);
	\end{scope}
	\end{tikzpicture}
	\vspace*{1.65mm}
	\caption{\label{fig:exist_centralized}centralized}
\end{subfigure}
\begin{subfigure}[b]{0.35\linewidth}
	\centering
	\begin{tikzpicture}[thick,node distance=1.3cm,>=latex]
	\begin{scope}
	  \def\fs{\small}  
	  \tikzstyle{block}=[circle,draw,fill=black!6,minimum height=2.1em]
	  \tikzstyle{loopstyle}=[loop,looseness=8]
	  \node [block] (N1){$1$};
	  \node [block, right of=N1] (N2) {$2$};
	  \draw [->] (N1) edge node[above] {\fs$0$} (N2); 
	\end{scope}
	\begin{scope}[shift={(0.08,1)}]
	  \tikzstyle{loopstyle}=[out=90+35,in=90-35,loop,looseness=6]
	  \tikzstyle{block}=[rectangle,draw]
	  \def \vgap {30pt};
	  \node (w1) at (0,0) {$w^1$};
	  \node [right of=w1] (w2) {$w^2$}; 
	  \node[block] (s12) at (0,\vgap) {$\{1,2\}$};
	  \node[block,right of=s12] (s2) {$\{2\}$};
	  \draw[->,dashedx] (w1) edge (s12);
	  \draw[->,dashedx] (w2) edge (s2);
	  \draw[->] (s12) edge [loopstyle] (s12);
	  \draw[->] (s2) edge [loopstyle] (s2);
	\end{scope}
	\end{tikzpicture}
	\vspace*{-2.2mm}
	\caption{\label{fig:exist_sparsity}sparsity}
\end{subfigure}
\begin{subfigure}[b]{0.3\linewidth}
	\centering
	\begin{tikzpicture}[thick,node distance=1.4cm,>=latex]
	\begin{scope}
	  \def\fs{\small}  
	  \tikzstyle{block}=[circle,draw,fill=black!6,minimum height=2.1em]
	  \tikzstyle{loopstyle}=[loop,looseness=8]
	  \node [block] (N1){$1$};
	  \node [block, right of=N1] (N2) {$2$};
	  \draw [->] (N1) edge [out=30,in= 150] node[above] {\fs$1$} (N2);
	  \draw [->] (N2) edge [out=-150,in=-30] node[below] {\fs$1$} (N1);
	\end{scope}
	\begin{scope}[shift={(0.1,1)}]
	  \tikzstyle{loopstyle}=[out=90+35,in=90-35,loop,looseness=6]
	  \tikzstyle{block}=[rectangle,draw]
	  \def \vgap {30pt}
	  \def \hgap {0.6cm}
	  \node (w1) at (0,0) {$w^1$};
	  \node (w2) at (2*\hgap,0) {$w^2$}; 
	  \node[block] (s1) at (0,\vgap) {$\{1\}$};
	  \node[block] (s2) at (2*\hgap,\vgap) {$\{2\}$};
	  \node[block] (s12) at (\hgap,2*\vgap) {$\{1,2\}$};
	  \draw[->,dashedx] (w1) edge (s1);
	  \draw[->,dashedx] (w2) edge (s2);
	  \draw[->] (s1) edge  (s12);
	  \draw[->] (s2) edge  (s12);
	  \draw[->] (s12) edge [loopstyle] (s12);
	\end{scope}
	\end{tikzpicture}
	\vspace*{-2mm}
	\caption{\label{fig:exist_delay}delay}
\end{subfigure}
\caption{\label{fig:existing_results}
Three simple special cases.}
\end{figure}

In this section, we explain how Theorem~\ref{thm:main} specializes to the existing results mentioned in Section~\ref{sec:intro}. Representative network and information graphs for these simple examples is show in Fig.~\ref{fig:existing_results}.

\if\MODE1\bigskip\else\fi
\begin{cor}[Centralized case]
If the network graph has a single node as in Fig.~\ref{fig:exist_centralized}, the solution reduces to the standard linear quadratic regulator. 
\end{cor}
\begin{pf}
The information graph consists
of a single node with a self-loop. Thus, \eqref{eq:X-recursion}
reduces to the classical Riccati recursion and \eqref{eq:ss_sol}
implies that $\zeta_t^{\{1\}} = x_t^1$. 
\if\MODE1\qed\fi
\end{pf}
\begin{cor}[Sparsity constraints \cite{shah10,swigart_thesis}]
If the network graph has $N$ nodes with no delayed
edges as in Fig.~\ref{fig:exist_sparsity}, then the optimal gains can
be computed from $N$ classical Riccati recursions, one for each node.  
\end{cor}
\begin{pf}
Since the graph is acyclic, if $i$ and $j$ are distinct nodes, then
the instantaneously reachable sets $s_0^i$ and $s_0^j$ are
distinct. Furthermore, since there are no delays $s_t^i=s_0^i$ for
$t\ge 0$. Thus, the information graph consists of $N$ self-loops. Therefore, the solution from Theorem~\ref{thm:main} reduces to $N$
decoupled LQR solutions. 
\if\MODE1\qed\fi
\end{pf}
\begin{cor}[Delay constraints \cite{lamperskidynamic2012}]
If the network graph is strongly connected and all edges have a one-timestep delay, then the optimal gains can be computed as algebraic functions of a single classical Riccati solution. 
\end{cor}
\begin{pf}
Since the graph is strongly connected, the entire set of nodes $\mathcal{V}$ is reachable from any node~$i$. Thus, for any node $r\in\mathcal{U}$, there is a directed path to $\mathcal{V}$ and
$\mathcal{V}\to \mathcal{V}$ is the only self-loop. The recursions \eqref{eq:X-recursion} imply that all gains can be computed as algebraic functions of $X_{0:T}^{\mathcal{V}}$, which is computed from a classical Riccati recursion.
\if\MODE1\qed\fi
\end{pf}

\section{Limitations}\label{sec:discussion2}

We now discuss selected topics exploring the limitations of our work and  directions for possible future research.

\if\MODE1\textbf{Output feedback. }\else\paragraph{Output feedback.}\fi
In output feedback problems, the decision-makers have access to noisy
measurements of states rather than the states themselves. Solutions
are known for two-player
sparsity~\cite{lessardoptimal2012,lessardtpof_big,lessard_nayyar_tpof}, linear chain
topologies~\cite{tanaka_triangular}, star topologies~\cite{broadcast},
and one-timestep
delays~\cite{kurtaranlinearquadraticgaussian1974,sandellsolution1974,yoshikawadynamic1975}. A delayed output-feedback LQG problem was also considered for teleoperation applications~\cite{kristalny2012decentralized}.

Despite all these specific examples, it is not clear how the present work can be extended to output feedback over a general graph with mixed delays. Difficulties arise because of the challenge in determining appropriate sufficient statistics for dynamic programming when the aggregate delay between any two nodes is at least $2$ timesteps
\cite{nayyaroptimal2011,varaiyadelayed1978,yoshikawaseparation1978}. In the present work, sufficient statistics are computed by projecting the state onto the mutually orthogonal label sets.
It is unclear if such a decomposition is possible for output-feedback
problems. 
We mention one promising exception~\cite{lessard_nayyar_tpof}, in which sufficient statistics are derived along with a solution to the  finite-horizon version of the two-player problem. 

\if\MODE1\textbf{Correlated noise. }\else\paragraph{Correlated noise.}\fi
We assume in Problem~\ref{probmain} that the noises injected into the various nodes
are independent. This fact is used to show that the $\zeta_t^s$ states are mutually independent, thus enabling a critical simplification of the value function used in the dynamic programming argument. If the noises are correlated, for example $\ee(w_t^1 w_t^{2\,\tp}) \ne 0$, our approach fails.
Such problems are still partially nested, as defined in Section~\ref{sec:proof_linearity}, so the optimal controller is again unique and linear.
However, noise correlations fundamentally change the nature of the optimal controller structure. This phenomenon was studied in~\cite{lessard_decoupled} for a problem with two players and decoupled dynamics. It was found that if each player has $n$ states, the optimal controller may have a number of states proportional to $n^2$ when noises are correlated.

\if\MODE1\textbf{Realizability. }\else\paragraph{Realizability.}\fi
In general, a causal linear time-invariant system may be equivalently
represented using either state-space or transfer functions. However,
the two representations are not equivalent when we impose an
underlying graph structure and associated sparsity for the state-space
matrices~\cite{lessard_realizability,elia}. Specifically, every
structured state-space realization that is stabilizable and detectable
corresponds to a structured transfer function, but the opposite is not
true \cite{lessard_realizability}. We avoid the issue of
realizability by 
 defining the problem in state
space form, and deriving a state space controller that satisfies the
sparsity and delay constraints by construction.
As discussed in Section~\ref{sec:infinte_horizon}, all internal states~$\zeta^s$ of this controller are stable by construction, so our realization is stabilizing. For graphs with pure sparsity (no delays), there is no loss in assuming a realizable plant, because non-realizable plants can
never be stabilized using structured controllers~\cite{lessard_realizability}. However, no analogous result is known for systems with delays, as discussed in~\cite{elia}. 

\section{Numerical example}\label{sec:example}

This section gives a numerical example for Theorem~\ref{thm:main} and
Corollary~\ref{cor:inf}.  
\begin{exmp}\label{ex:riccatiConvergence}
Consider the four-node system depicted in Fig.~\ref{fig:fiveNode_graphs}. We show the network graph with its associated messages from Theorem~\ref{thm:msg} and the information graph and noise partition diagram derived in Section~\ref{sec:main}.

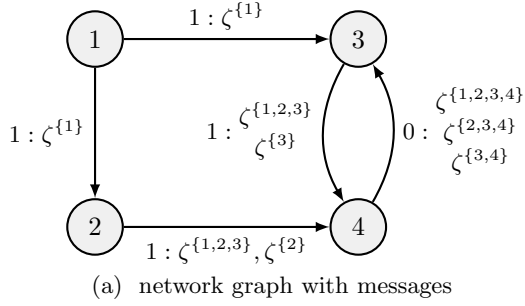
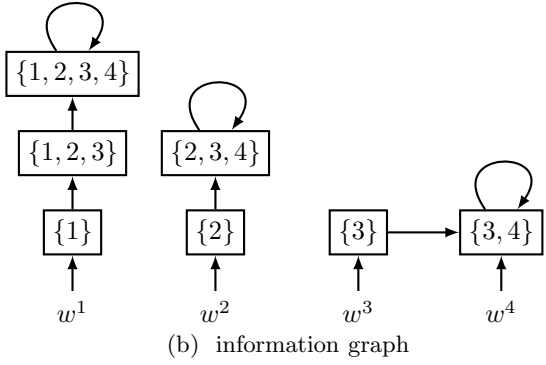
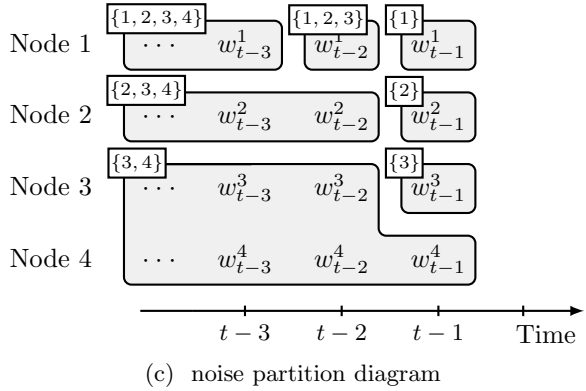
\begin{figure}[ht]
\centering
\begin{subfigure}[b]{\linewidth}
	\centering
	\begin{tikzpicture}[thick,node distance=3.5cm,>=latex]
		\def\fs{\small}  
		\tikzstyle{block}=[circle,draw, fill=black!6,minimum height=2.1em]
		\tikzstyle{loopstyle}=[loop,looseness=8]
		\node [block] (N1) {$1$};
		\node [block] (N2) at (0,-2.5) {$2$};
		\node [block,right of=N1] (N3) {$3$};
		\node [block,right of=N2] (N4) {$4$};
	        \draw [->] (N1) edge node[left] {\fs$1:\zeta^{\{1\}}$} (N2);
	        \draw [->] (N1) edge node[above] {\fs$1:\zeta^{\{1\}}$} (N3);
	        \draw [->] (N3) edge[out=-90-30,in=90+30] node[left] {\fs$1:
	\begin{matrix}\zeta^{\{1,2,3\}} \\ \zeta^{\{3\}}\end{matrix}$} (N4);
	        \draw [<-] (N3) edge[out=-90+30,in=90-30] node[right] {\fs$0:
	\begin{matrix}\zeta^{\{1,2,3,4\}} \\ \zeta^{\{2,3,4\}}
	  \\ \zeta^{\{3,4\}}
	\end{matrix}
	$} (N4);
	        \draw [->] (N2) edge node[below] {\fs$1:
	\zeta^{\{1,2,3\}}, \zeta^{\{2\}}
	$} (N4);
	\end{tikzpicture}
	\vspace{-2mm}
	\caption{\label{fig:fiveNode_messages}
	network graph with messages}
\end{subfigure}
\begin{subfigure}[b]{\linewidth}
	\centering
	\begin{tikzpicture}[thick,>=latex]
	  \tikzstyle{loopstyle}=[out=90+35,in=90-35,loop,looseness=6]
	  \tikzstyle{block}=[rectangle,draw]
	  \def \hgap {1.9cm}
	  \def \vgap {30pt}
	  \def \ha {0};
	  \def \hb {\hgap};
	  \def \hc {2*\hgap};
	  \def \hd {3*\hgap};
	  \def \he {4*\hgap};
	  \node (w1) at (\ha,0) {$w^1$};
	  \node (w2) at (\hb,0) {$w^2$};
	  \node (w3) at (\hc,0) {$w^3$};
	  \node (w4) at (\hd,0) {$w^4$};
	  \node[block] (S1) at (0,\vgap) {$\{1\}$};
	  \draw[->,dashedx] (w1) -- (S1);
	  \node[block] (S2) at (\hgap,\vgap) {$\{2\}$};
	  \draw[->,dashedx] (w2) -- (S2); 
	  \node[block] (S3) at (\hc,\vgap) {$\{3\}$};
	  \draw[->,dashedx] (w3) -- (S3);
	  \node[block] (S345) at ({\hd},\vgap) {$\{3,4\}$};
	  \draw[->,dashedx] (w4) -- (S345);
	  \node[block] (S123) at (0,2*\vgap) {$\{1,2,3\}$};
	  \draw[->] (S1) -- (S123);
	  \node[block] (S12345) at (0,3*\vgap) {$\{1,2,3,4\}$};
	  \draw[->] (S123) -- (S12345);
	  \draw[->] (S12345) edge [loopstyle] (S12345);
	  \node[block] (S2345) at (\hgap,2*\vgap) {$\{2,3,4\}$};
	  \draw[->] (S2) -- (S2345);
	  \draw[->] (S2345) edge [loopstyle] (S2345);
	  \draw[->] (S3) -- (S345);
	  \draw[->] (S345) edge [loopstyle] (S345);
	\end{tikzpicture}
	\vspace{-2mm}
	\caption{\label{fig:fiveNode_hierarchy}
	information graph}
\end{subfigure}
\begin{subfigure}[b]{\linewidth}
	\centering
	\begin{tikzpicture}[>=latex]
	\def\mdots{\,\,\cdot\cdot\cdot}
	\def\dy{0.22}
	\def\dax{1}
	\def\dt{0.07}
	\def\fs{\scriptsize}  
	\tikzstyle{boxes}=[draw,fill=black!6,rectangle,thick,rounded corners=1mm,inner
	sep=0ex]
	\tikzstyle{labels}=[draw,fill=white,rectangle,thick,inner xsep=0.3mm,inner
	ysep=0.7mm,anchor=north west]
	\matrix (m) [matrix of math nodes, row sep=0.9em, column sep=0.9em, ampersand
	replacement=\&]{
	\text{Node $1$}	\& \mdots \& w^1_{t-3} \& w^1_{t-2} \& w^1_{t-1} \&  \\
	\text{Node $2$}	\& \mdots \& w^2_{t-3} \& w^2_{t-2} \& w^2_{t-1} \&  \\
	\text{Node $3$}	\& \mdots \& w^3_{t-3} \& w^3_{t-2} \& w^3_{t-1} \&  \\
	\text{Node $4$}	\& \mdots \& w^4_{t-3} \& w^4_{t-2} \& w^4_{t-1} \& \phantom{w^4_t} \\[1mm]
	\& \& \text{\small $t-3$} \& \text{\small $t-2$} \& \text{\small $t-1$}\\ };
	\draw [thick,->] (m-4-3.north west)+(-0.2,-\dax) --
		node[pos=0.8,anchor=north west,inner ysep=5pt] {Time} +(5,-\dax);
	\draw [thick,dashedx] (m-4-3.north west)+(-0.2,-\dax) -- +(-0.9,-\dax);
	\draw [thick] (m-4-3.north)+(0,-\dax-\dt) -- +(0,-\dax+\dt);
	\draw [thick] (m-4-4.north)+(0,-\dax-\dt) -- +(0,-\dax+\dt);
	\draw [thick] (m-4-5.north)+(0,-\dax-\dt) -- +(0,-\dax+\dt);
	\draw [thick] (m-4-6.north)+(0,-\dax-\dt) -- +(0,-\dax+\dt);
	\tikzstyle{shadebox}=[rectangle,line width=0pt,rounded corners=1mm,inner
	sep=0ex]
	\pgfdeclarelayer{background}
	\pgfsetlayers{background,main}
	\begin{pgfonlayer}{background}
	  \node [boxes, fit=(m-1-2) (m-1-3)] (PL) {};
	  \node [boxes, fit=(m-1-4)] (QL) {};
	  \node [boxes, fit=(m-1-5)] {};
	  \node [boxes, fit=(m-2-2) (m-2-4)] (ML) {};
	  \node [boxes, fit=(m-2-5)] {};
	  \node [boxes, fit=(m-3-5)] {};
	  \node [shadebox,fit=(m-3-2) (m-3-4) (m-4-2) (m-4-4)] (NL) {};
	  \draw[rounded corners=1mm,thick, fill=black!6] 
	  (NL.north) -- (NL.north west) -- (NL.north east) -- (m-4-4.north east) --
	  (m-4-5.north east) -- (m-4-5.south east) -- (NL.south west) --
	  (NL.north west) -- (NL.north);
	\end{pgfonlayer}
	\path (m-1-5.north west)+(-\dy,\dy) node [labels] {\fs$\{1\}$};
	\path (m-1-4.north west)+(-\dy,\dy) node [labels] {\fs$\{1,2,3\}$};
	\path (PL.north west)+(-\dy,\dy) node [labels] {\fs$\{1,2,3,4\}$};
	\path (m-2-5.north west)+(-\dy,\dy) node [labels] {\fs$\{2\}$};
	\path (ML.north west)+(-\dy,\dy) node [labels] {\fs$\{2,3,4\}$};
	\path (m-3-5.north west)+(-\dy,\dy) node [labels] {\fs$\{3\}$};
	\path (NL.north west)+(-\dy,\dy) node [labels] {\fs$\{3,4\}$};
	\path (m-1-1.north)+(0,0.5) node {};
	\end{tikzpicture}
	\vspace{-2mm}
	\caption{\label{fig:fiveNode_noisesplit}
	noise partition diagram}	
\end{subfigure}
\caption{\label{fig:fiveNode_graphs}
Graphical representations for Example \ref{ex:riccatiConvergence}.}
\end{figure}

As a supplement to the discussion on infinite time horizon solutions in Section~\ref{sec:discussion2}, we now present a numerical simulation of this four-node example that shows convergence of the optimal control gains as the time horizon grows. We use the following time-invariant parameters,
 \begin{gather*}
 \if\MODE1\else\advance\arraycolsep-2pt\fi
 A = \bmat{
 3 & 0 & 0 & 0 \\
 2 & 3 & 0 & 0 \\
 1 & 2 & 2 & 1 \\
 0 & 1 & 3 & 2
 }
 \,
 B = \bmat{
 1 & 0 & 0 & 0 \\
 2 & 3 & 0 & 0 \\
 0 & 1 & 2 & 2 \\
 0 & 0 & 1 & 3
 }
 \,
 D = \bmat{0 & \infty & \infty & \infty \\
 1 & 0 & \infty & \infty \\
 1 & 1 & 0 & 0 \\
 2 & 1 & 1 & 0 }
 \\[1mm]
  \if\MODE1\else\advance\arraycolsep-2pt\fi
 Q=R = \bmat{
 8 & -1 & -1 & -1 \\
 -1 & 8 & -1 & -1 \\
 -1 & -1 & 8 & -1 \\
 -1 & -1 & -1 & 8
 }\,
 S = \bmat{
 -1 & -1 & -1 & -1 \\
 -1 & -1 & -1 & -1 \\
 -1 & -1 & -1 & -1 \\
 -1 & -1 & -1 & -1 
 }
 \end{gather*}
 \if\MODE1\else\vspace{-1mm}\fi
 and noise covariance $W = I$. Note that the $A$ and $B$ matrices have a sparsity pattern that conforms to the delay matrix $D$. As explained in Section~\ref{sec:prob_statement}, we must have $A^{ij}=0$ and $B^{ij}=0$ whenever $D^{ij} \ge 2$.

One can check that the systems matrices satisfy the conditions of Corollary~\ref{cor:inf}. It follows that the $X_t^r$ matrices, which are solutions to \eqref{eq:X-recursion}, should converge to steady-state values as we get farther from the terminal timestep. This fact is supported by the plot of Fig.~\ref{fig:riccatiConvergence}.
\if\MODE1\medskip\fi
 \begin{figure}[ht]
 \centering
 \begin{tikzpicture}
 \begin{axis}[
 	cycle list name=plain black,
 	xlabel={Time Step ($t$)}
 ]
 \pgfplotstableread{gainConvergence.txt}\table
  	\foreach \y in {1,2,...,7}
 		\addplot table[x index=0, y index=\y] from \table;
 \end{axis}
 \end{tikzpicture}
  \if\MODE1\else\vspace{-2mm}\fi
 \caption{\label{fig:riccatiConvergence} Plot of $\Trace(X_t^r)$ as a function of time for Example~\ref{ex:riccatiConvergence}. A curve is shown for each $r\in\mathcal{U}$.}
 \end{figure}
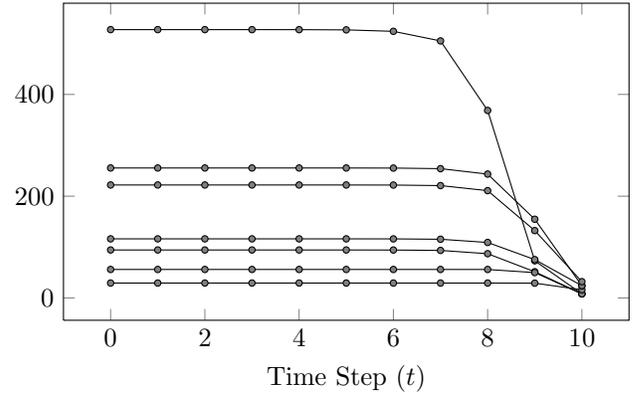
 \end{exmp}


\section{Proof of main results}\label{sec:mainproof}

This section contains proofs of Theorems~\ref{thm:main} and~\ref{thm:msg}.
The proof of Theorem~\ref{thm:main} formalizes and generalizes the strategy sketched in Section~\ref{sec:ex1sol}.

\subsection{Linearity}\label{sec:proof_linearity}

Linearity of the optimal policy follows from \emph{partial nestedness}, a concept first introduced by Ho and Chu in~\cite{hochu}. We state the main definition and result below.

\begin{defn}
A dynamical system \eqref{eq:ss_simplest} with information structure~\eqref{eq:info} is \emph{partially nested} if for every admissible policy $\gamma$, whenever $u_\tau^j$ affects $\Info_t^i$, then $\Info_{\tau}^j\subset \Info_t^i$.
\end{defn}

\begin{lem}[see \cite{hochu}]\label{lem:hochu}
Given a partially nested structure, the optimal control law that minimizes a quadratic cost of the form~\eqref{eq:cost_function} exists, is unique, and is linear.
\end{lem}

In other words, an information structure is partially nested if
whenever the decision of Player~$j$ affects the information used in
Player~$i$'s decision, then Player~$i$ must have access to all the
information available to Player~$j$. When this is the case, the
optimal policy is linear. Using partial nestedness, the following lemma shows that the optimal state and input may be expressed as linear functions of terms from the information sets $\Info_t^i$. 

\begin{lem} \label{lem:PN}
The information structure described in Problem~\ref{probmain} is
partially nested, so the optimal solution is linear and unique.
\end{lem}
\if\MODE1\vspace{-3mm}\fi
\begin{pf}
See Appendix~\ref{proofofHistoryLinearity}.
\end{pf}

\subsection{Disturbance-feedback representation}
 
As in Section~\ref{sec:ex1_disturbance}, the control inputs as
expressed as functions of the noise and initial conditions, in order
to exploit independence properties. 

Let $w_{-1}^i = x_0^i$, and define the \emph{noise information set} by
\begin{equation}\label{noiseinfoset}
\hat\Info_t^i = \bigl\{w_{k-1}^j:j\in \mathcal{V},\:\: 0\le k \le t-D^{ij}\bigr\}.
\end{equation}
\begin{lem}\label{lem:noiseInfo}
A collection of functions $\{\gamma^i_{0:T-1}\}_{i\in \mathcal{V}}$
satisfies the information constraint \eqref{eq:info} if and only if
there are functions $\{\hat\gamma^i_{0:T-1}\}_{i\in\mathcal{V}}$ such
that 
\begin{equation}\label{eq:noiseinfo}
u_t^i = \gamma_t^i( \Info_t^i ) = \hat\gamma_t^i(\hat\Info_t^i).
\end{equation}
\end{lem}
\begin{pf}
See Appendix~\ref{sec:noiseInfoPf}.
\end{pf}

As in Section~\ref{sec:ex1sol}, the paramterization in
\eqref{eq:noiseinfo} is an intermediate step that will enable us to
use a partition of the noise variables to decompose the inputs and
states into independent variables. The partition is defined in the
next lemma.

\begin{lem}\label{lem:infoset_labelset_equiv}
Consider an information graph $\hat G(\mathcal{U},\mathcal{F})$ and define the
corresponding \emph{label sets} $\left\{\Lab_{0:T}^s\right\}_{s\in\mathcal{U}}$ recursively by 
\begin{subequations}\label{eq:Lab1}
\begin{align}
\Lab_0^s &= \bigcup_{w^i\to s}\{x_0^i\} \\
\Lab_{t+1}^s &= \bigcup_{w_t^i\to s}\{w^i\} \cup \bigcup_{r\to s}\Lab_t^r.
\end{align}
\end{subequations}
The following properties of the label sets hold.
\begin{enumerate}[(i)]
\itemsep=-1mm
\item For every $t\ge 0$, the label sets are a partition of the noise history:
\begin{equation}\label{eq:labPartition}
\Lab_t^ r \cap \Lab_t^s = \emptyset
\text{ when $ r\neq s$, and }
\{w_{-1:t-1}\} = \bigcup_{s\in\mathcal{U}} \Lab_t^s.
\end{equation}
\item For all $i\in \mathcal{V}$,
\begin{equation}\label{harfharf}
\hat\Info_t^i = \bigcup_{s\ni i} \Lab_t^s.
\end{equation}
\end{enumerate}
\end{lem}
\if\MODE1\vspace{-3mm}\fi
\begin{pf}
See Appendix~\ref{proofofnoisepartition}. 
\end{pf}

Lemma~\ref{lem:PN} implies that the optimal solution is linear. When
policies are restricted to be linear, \eqref{harfharf} immediately
implies the following corollary.
\begin{cor}\label{cor:decomp}
A linear policy $\{\gamma_{0:T-1}^i\}_{i\in\mathcal{V}}$ is feasible
if and only if the inputs satisfy the following decomposition:
\begin{equation}\label{gen_inputdecomp}
u_t = \sum_{s\in\mathcal{U}} I^{\mathcal{V},s} \varphi_t^s,
\end{equation}
where $\varphi_t^s\in\Span(\Lab_t^s)$. 
\end{cor}

As before, the state can also be decomposed as a sum of terms from 
$\Span(\Lab_t^s)$. 

\begin{lem}\label{lem:stateDecomp}
Say that $\phi_t^s\in \Span(\Lab_t^s)$, and define $\zeta_t^s$ recursively by
\begin{subequations}\label{eq:zeta_both_repeat}
\begin{align} 
\zeta_0^s &= \sum_{w^i \to s} I^{s,\{i\}} x_0^i \label{eq:zetaInit_repeat}\\ \label{eq:zeta_repeat}
\zeta_{t+1}^s &= \sum_{ r \to s} \bigl(A_t^{s r}\zeta_t^r+B_t^{s r}\phi_t^r\bigr)+ \sum_{w^i \to s}I^{s,\{i\}}w_t^i .
\end{align}
\end{subequations}
Then $\zeta_t^s\in\Span(\Lab_t^s)$ and $x_t$ can be decomposed as
\begin{equation}\label{gen_statedecomp}
x_t = \sum_{s\in\mathcal{U}} I^{\mathcal{V},s} \zeta_t^s. 
\end{equation}
\end{lem}
\if\MODE1\vspace{-3mm}\fi
\begin{pf}
See Appendix~\ref{proofOfLemStateDecomp}.
\end{pf}
Note that~\eqref{eq:zeta_both_repeat} agrees with the
formula~\eqref{eq:ss_sol} given in the statement of
Theorem~\ref{thm:main}, provided that we set~$\phi_t^s = K_t^s
\zeta_t^s$. Corollary~\ref{cor:decomp} and
Lemma~\ref{lem:stateDecomp} imply that this policy is feasible. 

\begin{rem}\label{rem:estimation_interpretation}
We may interpret $\zeta_t^s$ and $\phi_t^s$ as conditional estimates of $x_t$ and $u_t$, respectively. Namely,
\[
\zeta_t^s = I^{s,\mathcal{V}} \ee( x_t \,|\, \Lab_t^s )
\quad\text{and}\quad
\phi_t^s = I^{s,\mathcal{V}} \ee( u_t \,|\, \Lab_t^s ).
\]
\end{rem}

\subsection{Optimality}\label{sec:proof_optimality}

We now prove the controller is optimal, and derive an expression for the corresponding minimal expected cost. Our proof uses a dynamic programming argument, and we optimize over \emph{policies} rather than \emph{actions}. Let $\gamma_t = \{\gamma_t^i\}_{i\in \mathcal{V}}$ be the set of policies at time~$t$. By~Lemma~\ref{lem:hochu}, we may assume the $\gamma_t^i$ are linear.  Define the cost-to-go
\begin{multline*}
V_t(\gamma_{0:t-1}) = \\
 \min_{\gamma_{t:T-1}} \ee^\gamma \Biggl(
\sum_{k=t}^{T-1} \bmat{x_k\\u_k}^\tp \bmat{Q_k & S_k \\ S_k^\tp & R_k} \bmat{x_k\\u_k}
+ x_T^\tp Q_f x_T^{\vphantom{\tp}} \Biggr),
\end{multline*}
where the expectation is taken with respect to the joint probability
measure on $(x_{t:T},u_{t:T-1})$ induced by the choice of $\gamma =
\gamma_{0:T-1}$. These functions are the minimum expected future cost
from time~$t$, given that the policies up to time $t-1$ have been
fixed. We allow $V_t$ to be a function of past policies, but
it turns out that $V_t$ will not depend on them
explicitly. 
By causality, we may iterate the minimizations and write a recursive
formulation for the cost-to-go, 
\begin{multline}\label{eq:costtogo_recursion}
V_t(\gamma_{0:t-1}) = \\
\min_{\gamma_t} \ee^\gamma \Biggl(
\bmat{x_t\\u_t}^\tp \bmat{Q_t & S_t \\ S_t^\tp & R_t} \bmat{x_t\\u_t}
+ V_{t+1}(\gamma_{0:t-1}, \gamma_t) \Biggr).
\end{multline}
Our objective is to find the optimal cost~\eqref{eq:cost_function}, which is simply~$V_0$. Consider the terminal timestep, and use the decomposition~\eqref{gen_statedecomp},
\begin{align*}
V_T(\gamma_{0:T-1}) = \ee^\gamma \bigl( x_T^\tp Q_f x_T \bigr)
= \ee^\gamma \sum_{s\in\mathcal{U}} (\zeta_T^s)^\tp Q_f^{ss} (\zeta_T^s).
\end{align*}
In the last step, we used the fact that the $\zeta_t^s$ coordinates
are independent. Note that $V_T$
depends on the policies up to time $T-1$ because the distribution of $\zeta_T^s$ depends on
past policies implicitly through~\eqref{eq:zeta}. We will prove by induction that the value function always has a similar quadratic form. Specifically, suppose that for some $t\ge 0$, we have
\[
V_{t+1}(\gamma_{0:t}) = \ee^\gamma \sum_{s\in\mathcal{U}} (\zeta_{t+1}^s)^\tp
X_{t+1}^s(\zeta_{t+1}^s) + c_{t+1},
\]
where $\{X_{t+1}^s\}_{s\in\mathcal{U}}$ is a set of matrices and
$c_{t+1}$ is a scalar. Now compute $V_t(\gamma_{0:t-1})$ using the
recursion~\eqref{eq:costtogo_recursion}. Substituting $\phi_t^s$ and
$\zeta_t^s$ from \eqref{gen_inputdecomp} and \eqref{gen_inputdecomp},
and using the independence of $\Lab_t^s$, we obtain
\begin{multline*}
V_t(\gamma_{0:t-1}) = \min_{\gamma_t} \ee^\gamma \Biggl(
\sum_{s\in\mathcal{U}} \bmat{\zeta_t^s\\\phi_t^s}^\tp
\bmat{Q^{ss}_t & S^{ss}_t \\ {S^{ss}_t}^\tp & R^{ss}_t}
\bmat{\zeta_t^s\\\phi_t^s} \\
+ (\zeta_{t+1}^s)^\tp X_{t+1}^s(\zeta_{t+1}^s) +
c_{t+1} \Biggr).
\end{multline*}
Substituting the state equations~\eqref{eq:zeta_repeat}, using the independence result
once more and rearranging terms, we obtain
\begin{equation}\label{eq:bellman_final}
V_t(\gamma_{0:t-1}) = \min_{\gamma_t} \ee^\gamma \sum_{r\in\mathcal{U}}
\bmat{\zeta^r_t \\ \phi^r_t }^\tp \Gamma^r_t \bmat{\zeta^r_t \\ \phi^r_t }
+ c_t,
\end{equation}
where $\Gamma^r_{0:T-1}$ and $c_{0:T-1}$ are given by:
\begin{align}
\Gamma_t^r  &= \bmat{Q_t^{rr} & S_t^{rr} \\ {S_t^{rr}}^\tp & R_t^{rr}} +
\bmat{A_t^{sr} & B_t^{sr}}^\tp X_{t+1}^s \bmat{A_t^{sr} & B_t^{sr}} \\
c_t &= c_{t+1} + \sum_{\substack{i \in \mathcal{V}\\ w^i \to s}}
\Trace\left(  (X_{t+1}^s)^{\{i\},\{i\}} W_t^i\right). \label{eq:cost_recursion2}
\end{align}
The terminal conditions are $\Gamma^r_T = Q_f^{rr}$ and $c_T=0$,
and~$s$ is the unique node in $\hat G(\mathcal{U},\mathcal{F})$ such that $r\to s$, see Proposition~\ref{prop:hierarchy_properties}. In~\eqref{eq:bellman_final}, the terms in the sum are independent, so they may be optimized separately. A lower bound on the cost-to-go is found by relaxing the information constraints and performing an unconstrained optimization over the actions $\phi_t = \{\phi_t^s\}_{s\in\mathcal{U}}$,
\begin{align*}
V_t(\gamma_{0:t-1})
&\geq  \ee^\gamma \sum_{r\in\mathcal{U}} \min_{\gamma_t} 
\bmat{\zeta^r_t \\ \phi^r_t }^\tp \Gamma^r_t \bmat{\zeta^r_t \\ \phi^r_t } + c_t
\\[-1mm]
& \geq \ee^\gamma  \sum_{r\in\mathcal{U}} \min_{\phi_t} 
\bmat{\zeta^r_t \\ \phi^r_t }^\tp \Gamma^r_t \bmat{\zeta^r_t \\ \phi^r_t } + c_t,
\end{align*}
where the first inequality follows from Fatou's lemma applied to~\eqref{eq:bellman_final}, and the second inequality follows from the relaxation mentioned above. Each minimization is a simple quadratic optimization, and the optimal cost and action are given by~\eqref{eq:X-recursion}. Substitution yields
\[
V_t(\gamma_{0:t-1}) \geq \ee^\gamma \sum_{s\in\mathcal{U}} (\zeta_{t}^s)^\tp
X_{t}^s(\zeta_{t}^s) + c_{t}.
\]
This lower-bound is in fact tight, because the optimal unconstrained actions are $\phi_t^s = K_t^s \zeta_t^s \in \Span\Lab_t^s$, which is precisely the admissible set for $\phi_t^s$. This completes the induction argument as well as the proof that the specified policy is optimal. The optimal cost is given by
\begin{align}
V_0 &= \ee \sum_{s\in\mathcal{U}} (\zeta_0^s)^\tp X_0^s (\zeta_0^s) + c_0 \nonumber \\
&= \ee \sum_{\substack{i\in \mathcal{V}\\w^i\to s}} (x_0^i) ^\tp (X_0^s)^{\{i\},\{i\}} (x_0^i) + c_0. \label{smurf}
\end{align}
where $c_0$ may be evaluated by starting with~$c_T=0$ and recursing backwards using~\eqref{eq:cost_recursion2}. Finally,~\eqref{smurf} evaluates to the desired expression~\eqref{eq:opt_cost} because $x_0^i \sim \mathcal{N}(0,\Sigma_0^i)$.
This completes the proof of Theorem~\ref{thm:main}.\qed

\subsection{Proof of Theorem~\ref{thm:msg}} \label{pf:msg}

Recall that there are no directed cycles in the network graph of delay
$0$. 
\begin{equation*}
(t,i) \prec (\tau,j) \text{ if } 
\bigl(
t<\tau
\bigr) \textrm{ or } 
\bigl(
t=\tau \textrm{ and } D^{ji} = 0
\bigr).
\end{equation*}
Recall that $D^{ji}$ implies that there is a directed path from $i$ to
$j$ of delay $0$. 
The proof will proceed by induction over this partial order. 

Let $\tilde\Info_t^i = \{x_t^i\}\cup \Mem_t^i \cup \bigcup_{j\nodelay i} \Msg_t^{ji} 
\cup \bigcup_{j\delay i} \Msg_{t-1}^{ji}$. If $t=0$ and $i$ has no
incoming delay-$0$ edges, then $\tilde\Info_0^i = \Info_0^i = \{x_0^i\}$. Thus \eqref{messagePassingSpan},
rewritten as $\zeta_t^s \in\Span(\tilde\Info_t^i)$, holds at $(t,i) =
(0,i)$. 

Fix $(t,i)$. Say that \eqref{messagePassingSpan} holds for all $(\tau,j)$. 
with $(\tau,j)\prec (t,i)$. Agent $i$ measures $x_t^i$ directly, by
assumption. If $j\nodelay i$, then $(t,j)\prec (t,i)$ implies that
$\Msg_t^{ji}$ could be computed and sent by agent $j$. If $t=0$, then
the local memory and incoming delay-$1$ messages are empty. If $t>0$,
then $(t-1,j)\prec (t,i)$ implies that the local memory $\Mem_t^i$ could be computed by agent $i$ at
time $t-1$ and that the messages $\Msg_{t-1}^{ji}$ could be computed
as well. Thus, $\tilde\Info_t^i$ can be computed.

Now it will be shown that \eqref{messagePassingSpan} holds at
$(t,i)$. 
Say that $i\in s$. If $t=0$, then $\zeta_0^s\ne 0$ implies that
$\zeta_0^s$ is a linear function of $x_0^j$, with $D^{ij}=0$. If
$j=i$, then $\zeta_0^s$ can be computed from the local measurement,
while if $j\ne i$, then $\zeta_0^s$ must have been contained in an
incoming message. 

Now say that $t>0$. 
First consider the case that $w^i\nrightarrow s$, so
that \eqref{eq:ss_update} reduces to 
\begin{equation*}
\zeta_{t}^s = \sum_{r\to s}\left(A_{t-1}^{sr}+B_{t-1}^{sr}K_{t-1}^r\right)\zeta_{t-1}^r.
\end{equation*}
If $i\notin r$, then $\zeta_{t-1}^r$ is contained in a delay-$1$
message $\Msg_{t-1}^{ji}$. So say that $i\in r$. If
$\zeta_{t-1}^r\in\Mem_{t}^i$, then it is already available to agent
$i$. 
Furthermore, if $\zeta_{t}^r\notin \Mem_t^i$, then it is contained in
some message $\Msg_{t-1}^{ji}$, where $j\nodelay i$. Since $i,j\in
r\subset s$, it follows that $\zeta_t^s$ is contained in message
$\Msg_t^{ji}$, so the equation above does not need to be computed.
In either case, it follows that $\zeta_t^s$ may be computed from the
combination of incoming messages and local memory. 

Now consider the case that $w^i\to s$. The subvector
$\left(\zeta_t^s\right)^{s\setminus \{i\}}$ can be computed as above
using
\begin{equation*}
\left(\zeta_t^s\right)^{s\setminus \{i\}} = I^{s\setminus\{i\},s}\sum_{r\to s}\left(A_{t-1}^{sr}+B_{t-1}^{sr}K_{t-1}^r\right)\zeta_{t-1}^r.
\end{equation*}
Since all vectors $\zeta_t^r$ with $i\in r\ne s$ can be computed as
above, the subvector $\left(\zeta_t^s\right)^i$ can be computed using
the state decomposition \eqref{gen_statedecomp}:
\begin{equation*}
\left(\zeta_t^s\right)^i = x_i - 
\sum_{
\substack{
r\ni i \\
r\ne s}}
\left(\zeta_t^r\right)^i.
\end{equation*}
Thus \eqref{messagePassingSpan} holds at $(t,i)$ and the proof is complete.
\qed

\section{Conclusion}\label{sec:conclusion}

This paper uses dynamic programming to derive optimal
policies for a general class of decentralized linear quadratic state
feedback problems. As noted in Section~\ref{sec:discussion1}, the
solution generalizes many existing works on decentralized
state-feedback control \cite{lamperskidynamic2012,shah10,swigart_thesis}. 

The key technique in the paper is the decomposition of available information based
on the \emph{information graph}. The graph is
used to specify both dynamics of the controller states, as well as the
structure of the Riccati difference equations required to compute the
solution. 

As discussed in Section~\ref{sec:discussion2}, many possible avenues for future research remain open. For example, some special cases with noisy measurements  or correlated noise have been solved, but extensions to general directed graphs with mixed sparsity and delays have yet to be found.

\section{Acknowledgments}

The first author thanks John Doyle for very helpful discussions.
The second author would like to thank Ashutosh Nayyar for some very helpful
discussions.

\if\MODE1
	\bibliographystyle{abbrv}
	\bibliography{sparsity_delays_auto}
\else
	{\small\bibliographystyle{abbrv}
	\bibliography{sparsity_delays_auto}}
\fi

\appendix

\section{Information set properties}
\label{pfDefEquiv}

We now prove some useful properties of information sets.

\begin{lem}\label{lem:infoDefEquiv}
The information sets~\eqref{infoset} may be expressed recursively as follows,
\begin{subequations}
\label{infoset_recursive}
\begin{align}
\Info_{-1}^i &= \emptyset \\
\Info_t^i &=  \bigl\{x_t^j: j\in\mathcal{V},\, D^{ij}=0\bigr\}\cup \bigcup_{\substack{j\in\mathcal{V}\\D^{ij}\le 1}}\Info_{t-1}^j
\quad\text{for }t \ge 0.
\end{align}
\end{subequations}
Similarly, the noise information sets from \eqref{noiseinfoset} can be
expressed recursively as
\begin{subequations}
\label{noiseinfoset_recursive}
\begin{align}
\hat\Info_{-1}^i &= \emptyset \\
\hat\Info_t^i &=  \bigl\{w_{t-1}^j: j\in\mathcal{V},\, D^{ij}=0\bigr\}\cup \bigcup_{\substack{j\in\mathcal{V}\\D^{ij}\le 1}}\hat\Info_{t-1}^j
\quad\text{for }t \ge 0.
\end{align}
\end{subequations}

Furthermore, suppose $i,j\in \mathcal{V}$ and $0\le k\le t$. 
The following are equivalent.
\begin{equation*}
\overbrace{\Info_k^j\subset \Info_t^i}^{(i)} 
\iff \overbrace{x_k^j\in \Info_t^i}^{(ii)}
\iff \overbrace{D^{ij}\le t-k}^{(iii)}
\end{equation*}
\end{lem}
\if\MODE1\vspace{-5mm}\fi
\begin{pf}
We first prove that {\it(i)--(iii)} are equivalent.
\begin{itemize}
\item {\it(i)$\implies$(ii)}: It is immediate from~\eqref{infoset} that
  $x_k^j\in \Info_k^j$. So it follows that $x_k^j \in \Info_t^i$. 
\item {\it(ii)$\implies$(iii)}: It follows from~\eqref{infoset} that
  if $x_k^j\in\Info_t^i$, then we must have $0\le k \le t-D^{ij}$. 
Therefore $D^{ij}\le t-k$ as required. 
\item {\it(iii)$\implies$(i)}:  By the triangle inequality, $D^{i\alpha} - D^{j\alpha} \le D^{ij}$ for any $\alpha\in\mathcal{V}$. Therefore, if $D^{ij} \le t-k$, then $k-D^{j\alpha} \le t-D^{i\alpha}$. So for any $\ell$ that satisfies $0 \le \ell \le k-D^{j\alpha}$, we must also have $0 \le \ell \le t-D^{i\alpha}$. It follows from~\eqref{infoset} that $\Info_k^j \subset \Info_t^i$.
\end{itemize}
We now derive the recursive expression for $\Info_t^i$. Start with~\eqref{infoset}, which we rewrite here for convenience,  
\[
\Info_t^i = \bigl\{ x_k^j : j\in\mathcal{V},\, 0 \le k \le t-D^{ij}
\bigr\}. 
\]
Partition into two cases; when $k=t$ (which implies $D^{ij}=0$), and when $k\ne t$. Then, partition further based on the value of $D^{ij}$.
\begin{align*}
\Info_t^i &= \bigl\{ x_t^j :  D^{ij}=0 \bigr\}
	\cup \bigl\{ x_k^j :  0\le k \le t-D^{ij},\, k\ne t \bigr\} \\
&= \bigl\{ x_t^j : j\in\mathcal{V},\, D^{ij}=0 \bigr\}
\\
	&\hspace{1cm}
\cup \bigl\{ x_k^j : j\in\mathcal{V},\, 0\le k \le
        t-1-D^{ij},\, D^{ij} = 0 \bigr\} \\
	&\hspace{1cm}
	\cup \bigl\{ x_k^j :j\in\mathcal{V},\,
 0\le k \le t-D^{ij},\, D^{ij}\ge 1 \bigr\}.
\end{align*}
In the last term, when $D^{ij}\ge 1$, it means that there is a path $j\to i$ with an aggregate delay of at least~1; so there exists an intermediate node $\ell \ne i$ where $D^{i\ell} = 1$ and $D^{\ell j} + 1 = D^{ij}$. Therefore,
\begin{align*}
\Info_t^i &= \bigl\{ x_t^j : j\in \mathcal{V},\,
D^{ij}=0 \bigr\}
\\&\hspace{1cm}
	\cup \bigl\{ x_k^j : j\in\mathcal{V},\, 0\le k \le
        t-1-D^{ij},\, D^{ij} = 0 \bigr\} \\
	&\hspace{1cm}
	\cup \bigcup_{\substack{\ell \ne i \\ D^{i\ell}=1}} 
\bigl\{ x_k^j : j\in\mathcal{V},\, 0\le k \le t-1-D^{\ell j}\bigr\} \\
&= \bigl\{ x_t^j :  D^{ij}=0 \bigr\}
	\cup \bigcup_{\substack{\ell \in \mathcal{V} \\ D^{i\ell}\le 1}} \bigl\{ x_k^j : 0\le k \le t-1-D^{\ell j}\bigr\} \\
&= \bigl\{ x_t^j : j\in\mathcal{V},\, D^{ij}=0 \bigr\}
	\cup \bigcup_{\substack{j \in \mathcal{V} \\ D^{ij}\le 1}}
	\Info_{t-1}^j,
\end{align*}
as required.

The proof of \eqref{noiseinfoset_recursive} is identical to the proof
of \eqref{infoset_recursive}, with $x_t^i$ replaced by $w_{t-1}^i$. 
\if\MODE1\qed\fi
\end{pf}

\section{Proof of Lemma~\ref{lem:PN}}
\label{proofofHistoryLinearity}

Suppose that $u_\tau^j$ affects $\Info_t^i$ in the simplest way possible; namely that $u_\tau^j$ affects $x_\sigma^\ell$ at a future timestep $\sigma > \tau$ via recursive applications of the state equations~\eqref{eq:ss_eqns}, and $x_\sigma^\ell \in \Info_t^i$. Then we have
\begin{align}
\label{eqq:pn1} u_\tau^j\text{ affects }x_{\sigma}^\ell &\implies D^{\ell j} \leq
\sigma-\tau \\
\label{eqq:pn2} x_{\sigma}^\ell \in \Info_t^i &\implies D^{i\ell} \leq t-\sigma.
\end{align}
Adding \eqref{eqq:pn1}--\eqref{eqq:pn2} together and using the triangle
inequality, we obtain $D^{ij} \leq t-\tau$. By Lemma~\ref{lem:infoDefEquiv}, it follows that $\Info_\tau^j \subset \Info_t^i$, as required. If $u_{\tau}^j$ affects $\Info_t^i$ via a more complicated path, apply the above argument to each consecutive pair of inputs along the path to obtain the chain of inclusions $\Info_{\tau}^j \subset \dots \subset \Info_t^i$.
Thus, the system is partially nested. 
\qed

\section{Proof of Lemma~\ref{lem:noiseInfo}}\label{sec:noiseInfoPf}
Say that inputs are generated by $u_t^i = \gamma_{t}^i(\Info_t^i)$. It
will be shown that all terms in $\Info_t^i$ can be expressed as
functions of the noise terms in $\hat\Info_t^i$, using knowledge of
the policy. Thus,
\eqref{eq:noiseinfo} will hold by an appropriate choice of
$\hat\gamma_t^i$. 

First note that $\Info_0^i=\hat\Info_0^i$. Now say that $t-1\ge 0$ and assume that the all
information sets $\Info_{t-1}^i$ can be computed from the
corresponding noise information sets $\hat\Info_{t-1}^i$. 
Say that $x_{\tau}^j\in \Info_t^i$. If $\tau<t$, then
\eqref{infoset_recursive} implies that $x_{\tau}^j\in \Info_{t-1}^k$
for some $k$ with $D^{ik}\le 1$. It follows that $x_{\tau}^j$ can be
computed from $\hat\Info_{t-1}^k$ which is contained in
$\hat\Info_t^i$, by \eqref{noiseinfoset_recursive}, from Lemma~\ref{lem:infoDefEquiv}. If $\tau=t$, then
$D^{ij}=0$ and
\eqref{eq:ss_eqns} implies that 
\begin{equation*}
x_t^j = \sum_{\substack{k\in \mathcal{V}\\D^{jk}\leq 1}}
\bigl( A^{jk}_{t-1} x_{t-1}^k + B^{jk}_{t-1} u_{t-1}^k \bigr) + w_{t-1}^j.
\end{equation*}
For each $k$ in the sum, $D^{ik}\le D^{ij}+D^{jk}\le 1$, so that
$x_{t-1}^i$ and $u_{t-1}^k$ are functions of terms from
$\hat\Info_{t-1}^k\subset\hat\Info_{t}^i$. Since
$w_{t-1}^j\in\hat\Info_t^i$, it follows that $x_t^j$ is computable
from $\hat\Info_t^i$.

Conversely, if the control actions satisfy $u_t^i =
\hat\gamma_t^i(\hat\Info_t^i)$ for some functions $\gamma_t^i$, an
analogous argument shows that all noise terms in $\hat\Info_t^i$ can
be deduced from $\Info_t^i$. 
\qed
\section{Proof of Lemma~\ref{lem:infoset_labelset_equiv}}
\label{proofofnoisepartition}
\emph{Part (i).} 
We proceed by induction. Set $\Hist_t = \{w_{-1:t-1}\}$. 
 At $t=0$, we have $\Hist_0=\{x_0^1,\ldots,x_0^N\}$. Since each~$w^i$ points to exactly one element $s\in\mathcal{U}$, it is clear from~\eqref{eq:Lab1} that $\{\Lab_0^s\}_{s\in\mathcal{U}}$ partitions $\Hist_0$. Now suppose that $\{\Lab_t^s\}_{s\in\mathcal{U}}$ partitions $\Hist_t$ for some $t\ge 0$. By Proposition~\ref{prop:hierarchy_properties}, for each $ r\in\mathcal{U}$ there exists a unique $s\in\mathcal{U}$ such that $ r\to s$. Therefore each element $w_k^i\in \Hist_t$ is contained in exactly one label set $\Lab_{t+1}^s$. It follows from~\eqref{eq:Lab1} that $\{\Lab_{t+1}^s\}_{s\in\mathcal{U}}$ must partition $\Hist_{t+1}$ and the proof is complete.

\emph{Part (ii).} Again, by induction. At $t=0$,
\begin{align*}
\Info_0^i
&= \{x_0^j: D^{ij}=0 \} \\
&= \{x_0^j: s_0^j \ni i \} \\
&= \{x_0^j: w^j \to s, s \ni i \} \\
&= \bigcup_{s\ni i} \Lab_0^s.
\end{align*}
So the identity holds at $t=0$. Now suppose it holds for some $t\ge
0$. By Lemma~\ref{lem:infoDefEquiv}, and the inductive hypothesis,
\begin{align*}\label{eq:ttmp}
\hat\Info_{t+1}^i &= \{ w_{t}^j: j \in \mathcal{V},\, D^{ij} = 0\}
	\cup \bigcup_{\substack{j\in\mathcal{V}\\D^{ij}\le 1}}
        \hat\Info_t^j \\
&= \{ w_{t}^j: j \in \mathcal{V},\, D^{ij} = 0\}
	\cup \bigcup_{\substack{j\in\mathcal{V}\\D^{ij}\le 1}}
        \bigcup_{j\ni r}\Lab_t^r.
\end{align*}
Note that $D^{ij}=0$ if and only if $w^j\to s$ for some $s\ni
i$. Furthermore, $D^{ij}\le 1$ and $j \in r$ implies that $r\to s$ for some $s\ni
i$. Conversely, if $s\ni i$, all sets $r$ with $r\to s$ must contain a
node $j$ with $D^{ij}\le 1$. Thus, \eqref{harfharf} holds.
\qed

\section{Proof of Lemma \ref{lem:stateDecomp}}
\label{proofOfLemStateDecomp}
Suppose $\zeta_t^s$ satisfies
\eqref{eq:zeta_both_repeat}. The recursive formulation of the label sets~\eqref{eq:Lab1} implies that $\zeta_t^s \in\Span \Lab_t^s$ for all $t\ge 0$ and all $s\in\mathcal{U}$. All that remains to be shown is that $\zeta_t^s$ indeed provides decomposition of $x_t$ as in \eqref{gen_statedecomp}. The decomposition of $x_t$ is satisfied at $t=0$, since for all $i\in \mathcal{V}$ there is a unique $s\in\mathcal{U}$ with $w_i\to s$. Now assume inductively that \eqref{gen_statedecomp} holds for some $t\ge 0$. Therefore,
\begin{align*}
\MoveEqLeft[1]
\sum_{s\in\mathcal{U}} I^{\mathcal{V},s} \zeta_{t+1}^s
\\
&= \sum_{s\in\mathcal{U}} I^{\mathcal{V},s}\left( \sum_{ r \to s} \bigl(A_t^{s r}\zeta_t^ r+B_t^{s
r}\phi_t^ r\bigr)+ 
\sum_{w^i \to s}I^{s,\{i\}}w_t^i \right) \\
&= \sum_{s\in\mathcal{U}} \left( \sum_{ r \to s} \bigl(A_t I^{\mathcal{V}, r}\zeta_t^ r+B_t I^{\mathcal{V},
r}\phi_t^ r\bigr)+ 
\sum_{w^i \to s}I^{\mathcal{V},\{i\}}w_t^i \right) \\
&= \sum_{s\in\mathcal{U}} \bigl(A_t I^{\mathcal{V},s}\zeta_t^s+B_t I^{\mathcal{V},s}\phi_t^s\bigr) +w_t^i  \\
&= A_tx_t + B_tu_t + w_t \\
&= x_{t+1},
\end{align*}
where we have substituted the dynamics~\eqref{eq:zeta_repeat} in the first step, and the induction hypothesis in the fourth step. In the second step, we took advantage of the particular sparsity structures of $A$ and $B$, that imply
\[
A_tI^{\mathcal{V}, r} = I^{\mathcal{V},s}A_t^{s r}
\qquad\text{and}\qquad
B_tI^{\mathcal{V}, r} = I^{\mathcal{V},s}B_t^{s r}.
\]
It follows that~\eqref{gen_statedecomp} holds for $t+1$.\qed

\end{document}